\documentclass[%
 reprint,
 amsmath,amssymb,
 aps,twocolumn
]{revtex4}

\usepackage{natbib}
\usepackage{aas_macros}
\usepackage{graphicx}
\usepackage{dcolumn}
\usepackage{bm}

\usepackage{float}

\usepackage{makecell}
\usepackage{ulem}

\newcommand\ba{\begin{eqnarray}}
\newcommand\ea{\end{eqnarray}}

\usepackage{color}

\begin{document}


\title{Probing initial isocurvature perturbation with 21cm one-point statistics }

\author{Zhenfei Qin}
 \affiliation{%
South-Western Institute for Astronomy Research (SWIFAR), Yunnan University, Kunming, Yunnan 650500, People's Republic of China
\\
Yunnan Key Laboratory of Survey Science, Kunming, Yunnan 650500, People's Republic of China
}%

 \email{zhenfei@mail.ynu.edu.cn}

\author{Hayato Shimabukuro}
 \affiliation{South-Western Institute for Astronomy Research (SWIFAR), Yunnan University, Kunming, Yunnan 650500, People's Republic of China
\\
Yunnan Key Laboratory of Survey Science, Kunming, Yunnan 650500, People's Republic of China\\
Graduate School of Science, Division of Particle and Astrophysical Science, Nagoya University, Chikusa-Ku, Nagoya, 464-8602, Japan}
  
 \email{shimabukuro@ynu.edu.cn}

\date{\today}

\begin{abstract}

Isocurvature perturbations—expected from multi-field inflation models—can leave unique signatures in the early Universe, but remain weakly constrained, especially on small scales. In this work, we investigate the constraining power of one-point statistics (variance and skewness) of the 21cm brightness temperature during Cosmic Dawn and the Epoch of Reionization, using semi-numerical simulations from 21cmFAST. We model both adiabatic and cold dark matter isocurvature modes, exploring their impact on the matter power spectrum, the timing of structure formation, and the evolution of neutral hydrogen. By varying astrophysical parameters as well as the isocurvature fraction and spectral index, we quantify their respective effects on the 21cm power spectrum and on one-point statistics, using the former as a physical diagnostic and the latter as the basis of our final forecast. Our results show that while variance is highly sensitive to the timing of cosmic events and provides tight constraints on isocurvature parameters, skewness is more strongly affected by astrophysical uncertainties and observational noise. Incorporating realistic instrumental noise based on SKA configurations, we perform a Fisher analysis using only the redshift evolution of the one-point statistics, and demonstrate that the isocurvature fraction can be constrained down to the percent level, though a strong degeneracy with the spectral index remains. We discuss the importance of complementary probes, such as the 21cm forest and galaxy surveys, to break these parameter degeneracies. Our findings highlight the power of 21cm one-point statistics as robust and independent tools for probing early-Universe physics beyond what is accessible with traditional power spectrum analyses.

\end{abstract}

\maketitle


\section{Introduction}

Understanding the early Universe is a fundamental goal of modern cosmology. The formation and evolution of cosmic structures are believed to be seeded by primordial fluctuations, which are imprinted on the cosmic microwave background (CMB). Observations of the CMB anisotropies, particularly those from the Planck satellite \citep{2020A&A...641A..10P,2011ApJS..192...18K}, have provided precise constraints on these primordial fluctuations, revealing that the primordial power spectrum is predominantly adiabatic (adi). In adiabatic perturbations, the relative number densities of different particle species remain constant, leading to fluctuations in the overall energy density without altering the composition of the Universe. However, this standard scenario does not exclude the possibility of additional components, such as isocurvature (iso) perturbations \citep{PhysRevD.62.083508,2009PhRvD..80l3516V}, which represent variations in the composition of the Universe rather than fluctuations in its overall density. Isocurvature perturbations can arise from mechanisms like multi-field inflation or cosmic defects and could have played an important role during the early stages of the Universe. These models predict the isocurvature fluctuation spectrum is blue-tilted \citep[e.g.][]{Kasuya_2009,Chung_2015,Chung_2018,Chung_2022,Afshordi_2003,Kashlinsky_2016,Gong_2017,Gong_2018,Mena_2019,Tashiro_2021}. Constraining these perturbations further is essential for improving our understanding of the inflationary era and the physics of the early Universe \citep{2000PhRvD..61l3507L,2000Adiabatic}.

One promising avenue for exploring isocurvature perturbations is the 21cm hydrogen line, which traces the neutral hydrogen distribution in the Intergalactic medium (IGM) throughout the Cosmic Dawn (CD) and the Epoch of Reionization (EoR). The 21cm line corresponds to the hyperfine transition of neutral hydrogen atoms and serves as a powerful tool for mapping the IGM in three dimensions. The 21cm signal provides a powerful probe of cosmology and astrophysics at the CD/EoR \citep[e.g.][]{2012RPPh...75h6901P,2016arXiv160301961H,2016PASJ...68R...2Y,2023PASJ...75S...1S}. The 21cm line can offer unique insights into the nature of the initial perturbations that seeded structure formation. Recently, \citet{2022PhRvD.105h3523M} have shown that the global 21cm signal can be used to constrain isocurvature perturbations, highlighting the potential of this observational method.

In our work, we build on this idea by focusing on the one-point statistics of the 21cm line signal, particularly its variance and skewness. By analyzing the distribution of 21cm brightness temperatures at individual points, we can capture non-Gaussian features via higher-order statistics\citep[e.g.][]{2016MNRAS.458.3003S,2016PASJ...68...61K,2017MNRAS.468.1542S}. We expect that these one-point statistics provide us with information on the isocurvature modes, which is complementary to power spectrum analysis, and offer an alternative means to distinguish between adiabatic and isocurvature contributions. One-point statistics of the 21cm signal can reveal features in the distribution of matter in addition to other methods\citep[e.g.][]{2014MNRAS.443.3090W,2015MNRAS.451..467S,2015MNRAS.454.1416W}. As multiple telescopes and experiments are either already observing or preparing to observe the 21cm signal from the CD and EoR—such as the Hydrogen Epoch of Reionization Array (HERA) \citep{2022ApJ...924...51A}, the Square Kilometre Array (SKA) \citep{2015aska.confE...1K}, and the Low-Frequency Array (LOFAR) \citep{2013A&A...556A...2V}—our method could play a crucial role in constraining isocurvature perturbations and, by extension, improving our understanding of the fundamental physics that governed the early Universe. This work provides a systematic forecast of cold dark matter
isocurvature constraints from 21cm one-point statistics under
realistic SKA-like noise conditions, using the 21cm power
spectrum as a physical diagnostic rather than as a Fisher
forecast observable.

The structure of this paper is organized as follows. In Section \ref{CDM isocurvature perturbations and matter power spectrum}, we summarize how to define and calculate the matter power spectrum with cold dark matter (CDM) isocurvature perturbations. We detail the modifications to the standard cosmological perturbation theory required to include isocurvature modes and discuss their impact on the matter power spectrum. The 21cm power spectrum, one-point statistics, thermal noise, and Fisher matrix are calculated in Section \ref{Cosmological 21cm signal}. Here, we outline the simulation setup, the statistical techniques employed, and the assumptions made regarding the instrumental configurations of upcoming 21cm experiments. In Section \ref{Sec3}, we present our results, which are followed by a summary and conclusion in Section \ref{Sec4}. Throughout, we adopt a standard $\Lambda$CDM cosmology with $h=0.673$, $\Omega_{\rm m}=0.316$, $\Omega_{\Lambda}=0.684$, $\Omega_{\rm b}=0.049$, $\sigma_8=0.811$, and $n_{\rm s}=0.96$, as constrained by Planck \citep{2020A&A...641A..10P}.


\section{The impacts of isocurvature perturbations on the structure formation}
\label{CDM isocurvature perturbations and matter power spectrum}


We assume that the power spectrum of the initial isocurvature perturbation is similar in form to that of the initial adiabatic perturbation, described by the following equations:

\begin{equation}
\mathcal{P}_\zeta (k) = A_\mathrm{s}^\mathrm{adi} \left(\frac{k}{k_*}\right)^{n_\mathrm{s}^\mathrm{adi}-1},
\label{eqn:inipsadia}
\end{equation}

\begin{equation}
\mathcal{P}_{S_\mathrm{CDM}} (k) = A^\mathrm{iso} \left(\frac{k}{k_*}\right)^{n^\mathrm{iso}-1},
\label{eqn:inipsiso}
\end{equation}

where $A_\mathrm{s}^\mathrm{adi}$ and $A^\mathrm{iso}$ denote the amplitudes of the initial adiabatic and isocurvature perturbation power spectra, respectively. The spectral indices, $n_\mathrm{s}^\mathrm{adi}$ and $n^\mathrm{iso}$, characterize the scale dependence of these spectra. The pivot scale $k_*$ is conventionally set to $0.05\,\mathrm{Mpc}^{-1}$, consistent with standard practices in CMB data analysis.

The transfer functions for adiabatic and CDM isocurvature perturbations are derived from linear perturbation theory and encapsulate the evolution of these perturbations across different scales during the radiation- and matter-dominated epochs. These transfer functions, denoted as $T_\mathrm{adi}(k)$ and $T_\mathrm{iso}(k)$, have been extensively studied in the literature \citep{1986ApJ...304...15B,1995ApJS..100..281S}. $T_\mathrm{adi}(k)$ describes the evolution of adiabatic modes, while $T_\mathrm{iso}(k)$ represents the evolution of isocurvature modes, reflecting their different physical origins and dynamical evolution.

Assuming that adiabatic and isocurvature perturbations are uncorrelated, the total matter power spectrum can be expressed as the sum of their contributions \citep{2022PhRvD.105h3523M}:

\begin{align}
P_\mathrm{m}(k) &= \mathcal{P}_\zeta(k) T_\mathrm{adi}^2(k) 
+ \mathcal{P}_{S_\mathrm{CDM}}(k) T_\mathrm{iso}^2(k) \nonumber \\
&= A_\mathrm{s}^\mathrm{adi} 
\left(\frac{k}{k_*}\right)^{n_\mathrm{s}^\mathrm{adi}-1} \notag \\
&\quad \times \left[T_\mathrm{adi}^2(k) 
+ r_\mathrm{CDM} \left(\frac{k}{k_*}\right)^{n^\mathrm{iso}-n_\mathrm{s}^\mathrm{adi}} 
T_\mathrm{iso}^2(k)\right],
\label{eq:power_tot}
\end{align}

where $r_{\rm CDM}$ is defined at the pivot scale
$k_\ast=0.05\,{\rm Mpc}^{-1}$ through the ratio of the primordial isocurvature to adiabatic power,
\[
r_{\rm CDM}\equiv
\frac{\mathcal{P}_{\rm iso}(k_\ast)}
{\mathcal{P}_{\rm adi}(k_\ast)}
=
\frac{A_{\rm iso}}{A_{\rm adi}} .
\]
At a general scale $k$, the relative isocurvature contribution is then
\[
\frac{\mathcal{P}_{\rm iso}(k)}
{\mathcal{P}_{\rm adi}(k)}
=
r_{\rm CDM}
\left(\frac{k}{k_\ast}\right)^{n_{\rm iso}-n_{\rm adi}} .
\]
Thus, for blue-tilted isocurvature spectra, the small-scale isocurvature contribution can be significantly enhanced. This formulation (Eq.~\ref{eq:power_tot}) highlights how adiabatic and isocurvature perturbations contribute to the overall matter power spectrum, with each term modulated by its respective transfer function and initial power spectrum. The choice of parameters, such as $r_\mathrm{CDM}$ and $n^\mathrm{iso}$, and their constraints play a crucial role in interpreting cosmological observations \citep{2016A&A...594A..13P,2016A&A...594A..17P}. While current CMB observations, such as those from Planck, constrain the isocurvature fraction to below approximately $1\%$ at large scales(\citep[e.g.][]{2020A&A...641A..10P,buckley2025generalconstraintsisocurvaturecmb}), these constraints are primarily applicable to large-scale (low-$k$) modes. In scenarios where the isocurvature spectrum is blue-tilted (i.e., with a large spectral index), the contribution at small scales can be significantly enhanced, and the CMB constraints become less stringent or more model-dependent in this regime. Therefore, investigating larger values of the isocurvature fraction (e.g., $5\%$ or $10\%$) in the context of 21cm statistics (explained later) is justified, both as a theoretical exploration and to evaluate the sensitivity of 21cm observables to such perturbations.

For the uncorrelated CDI case with free $n_{\rm iso}$, the Planck constraints should be interpreted as joint constraints in the $(r_{\rm CDM},n_{\rm iso})$ plane rather than as two independent one-dimensional bounds. Since the published Planck limits are usually quoted as upper limits on the isocurvature fraction at several reference scales, we use these limits to construct a conservative CMB-motivated upper envelope $r_{\max}(n_{\rm iso})$ constrained by Planck \citep{2020A&A...641A..10P}, shown in Appendix Fig.~\ref{fig:planck_envelope}. The shaded region in that figure indicates the parameter space below this conservative envelope. Our fiducial point, $(r_{\rm CDM},n_{\rm iso})=(0.05,2.5)$, marked by the red star, lies well inside this region. We therefore use it as a representative benchmark that remains compatible with the conservative CMB-motivated envelope while still producing a visible response in the 21cm observables.

\begin{figure}
    \includegraphics[width=0.85\hsize]{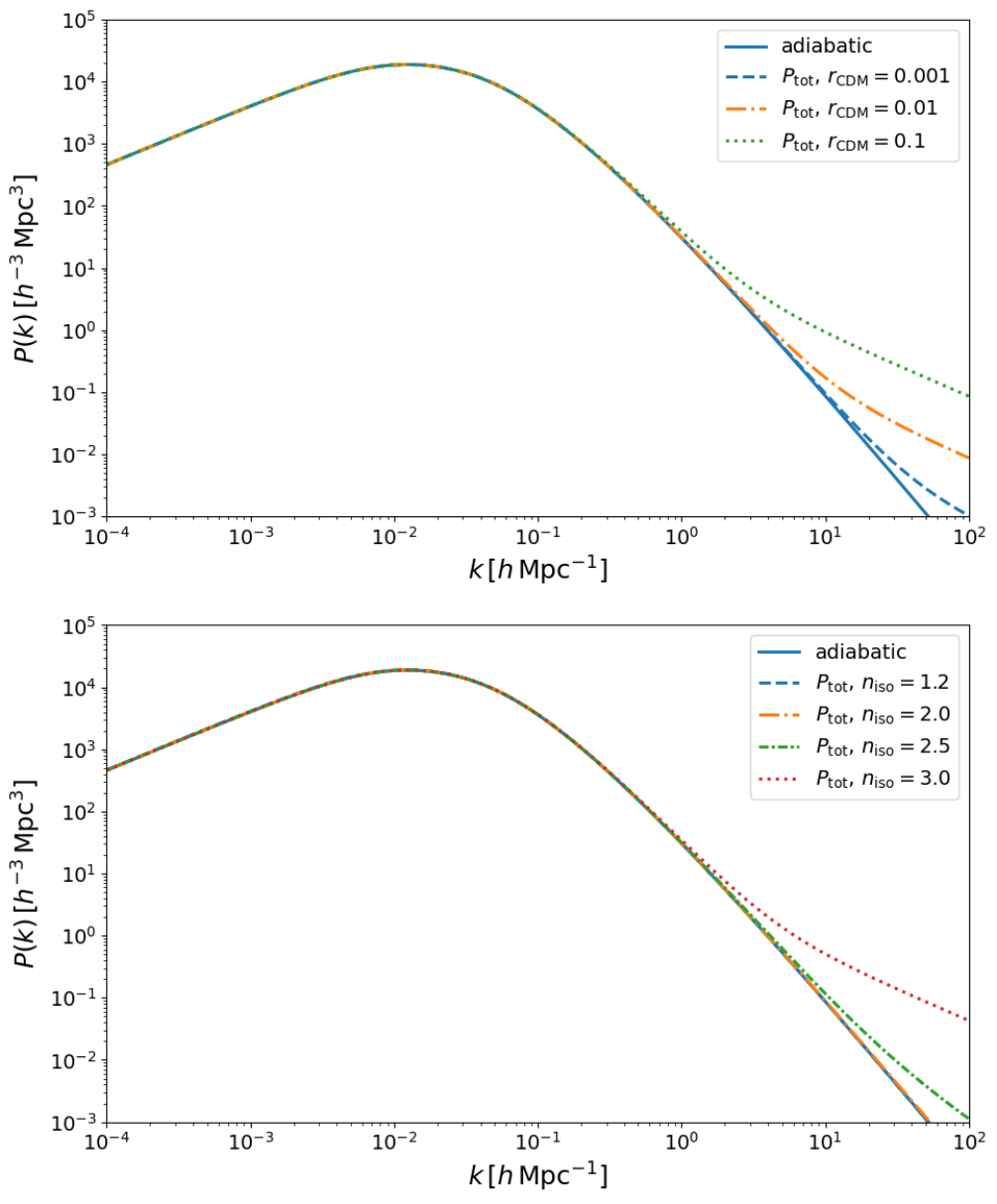}
    \caption{Total matter power spectrum, including both adiabatic and CDM isocurvature contributions, for different isocurvature parameters. Top: $P_{\rm tot}(k)$ for $r_{\rm CDM}=0.001$, $0.01$, and $0.1$ at fixed $n_{\rm iso}=3.0$. Bottom: $P_{\rm tot}(k)$ for $n_{\rm iso}=1.2$, $2.0$, $2.5$, and $3.0$ at fixed $r_{\rm CDM}=0.05$. The blue solid line denotes the purely adiabatic case.}
\label{fig:all_matter_ps}
\end{figure}

In Fig.~\ref{fig:all_matter_ps}, we illustrate the total matter power spectrum under varying values of $r_\mathrm{CDM}$ and $n^\mathrm{iso}$. The blue solid line denotes the purely adiabatic case, while the colored curves show the corresponding total spectrum
\[
P_{\rm tot}(k)=P_{\rm adi}(k)+P_{\rm iso}(k).
\]
The top panel shows the matter power spectrum for varying amplitudes of isocurvature perturbations ($r_\mathrm{CDM} = 0.1$, $0.01$, and $0.001$) while fixing $n^\mathrm{iso} = 3.0$. Increasing $r_{\rm CDM}$ raises the overall contribution of the isocurvature component, while the fact that the enhancement is most visible on small scales is a consequence of the adopted blue tilt. The bottom panel shows the impact of varying the spectral index $n^\mathrm{iso}$ ($n^\mathrm{iso} = 1.2$, $2.0$, $2.5$, and $3.0$) while fixing $r_\mathrm{CDM} = 0.05$. Changes in $n^\mathrm{iso}$ affect the slope of the power spectrum, with higher $n^\mathrm{iso}$ values producing more power on small scales. This highlights the role of the spectral index in shaping the distribution of matter across different scales. The figure also shows that the total matter power spectrum remains very close to the adiabatic case when $n^{\rm iso}$ is small.

These results underscore the distinct roles of $r_\mathrm{CDM}$ and $n^\mathrm{iso}$ in shaping the matter power spectrum when the spectrum is sufficiently blue. Enhanced isocurvature perturbations can accelerate the formation of small-scale structures, leading to earlier formation of the first stars and galaxies. Such changes could leave detectable imprints in the 21cm line signal, providing a potential avenue for probing the influence of isocurvature perturbations during the early Universe.

\begin{figure}
    
\includegraphics[width=1.0\hsize]{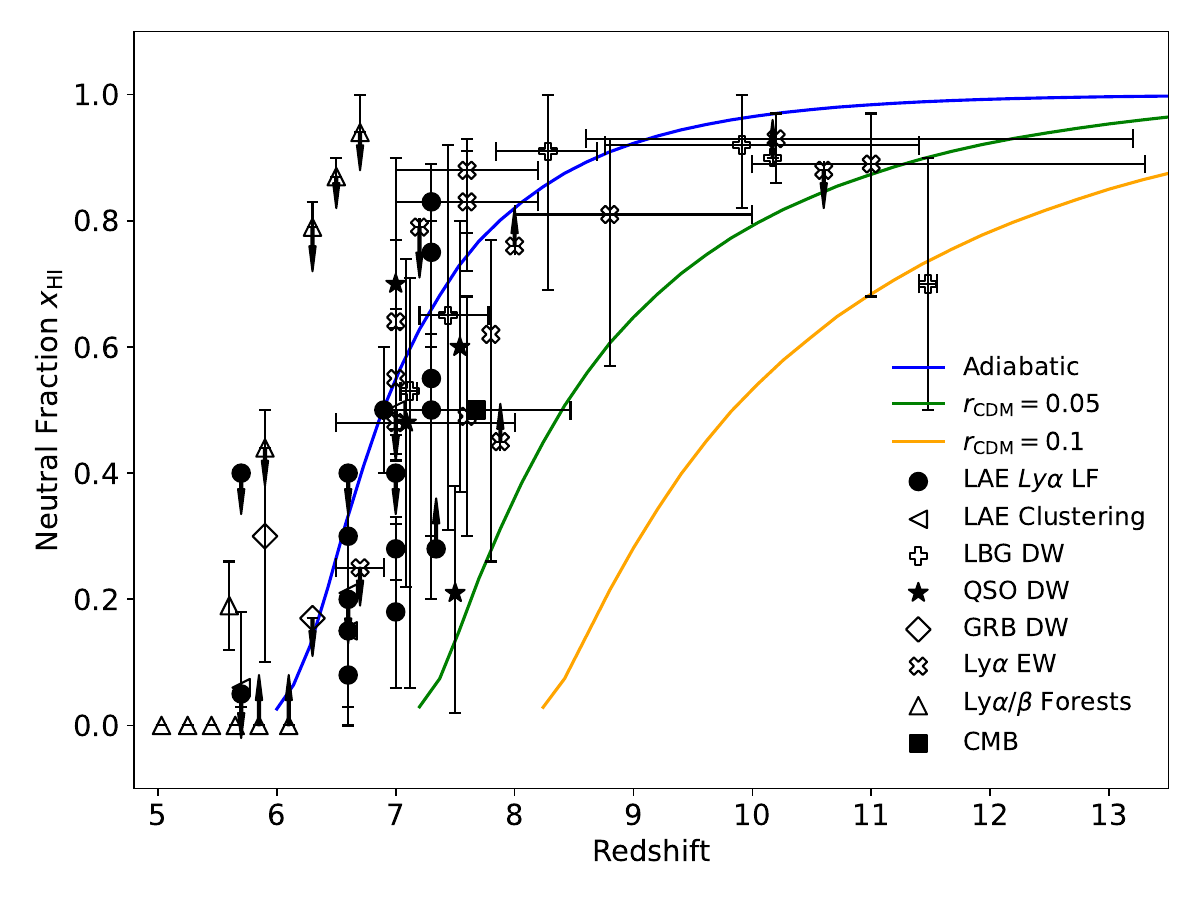}
\centering
    \caption{The theoretically predicted history of the neutral fraction for isocurvature scenarios and observational constraints taken from \citep{2024arXiv241115495U}.
    The lines express the evolution of neutral HI fraction for $r_\mathrm{CDM}$ =0, 0.05, 0.1, respectively. Here we fix $n^\mathrm{iso}=2.5$. We also plot some observational constraints to compare the neutral HI fraction of different isocurvature models. Filled circles: The LAE Lyman-$\alpha$ LF \cite{Ouchi_2010,Konno_2014,Zheng_2017,2018PASJ...70...55I,Morales_2021,Goto_2021,Ning_2022,umeda2024silverrushxivlyaluminosity}, left-tipped triangles: LAE Clustering \citep{2015MNRAS.453.1843S,2018PASJ...70S..13O,umeda2024silverrushxivlyaluminosity}, pluses: Lyman-$\alpha$ damping wing measurements of LBGs \citep{Umeda_2024,2023NatAs...7..622C,hsiao2024jwstnirspecspectroscopytriplylensed} , 
    filled pentagons: damping wing measurements of QSOs \citep{Davies_2018,2019MNRAS.484.5094G,Wang_2020}, 
    diamonds: damping wing measurements of GRBs \citep{2006PASJ...58..485T,2014PASJ...66...63T},
    X marks: Lyman-$\alpha$ equivalent width distributions \citep{Hoag_2019},
    filled square: CMB Thomson optical depth \citep{refId0}, 
    up-pointing triangles: the Gunn-Peterson trough of QSOs \citep{Fan_2006}.}    
\label{fig:nf}
\end{figure}

In Fig. \ref{fig:nf}, we show the evolution of the neutral hydrogen fraction ($x_\mathrm{HI}$) in the IGM as a function of redshift for various values of $r_\mathrm{CDM}$. These constraints are primarily derived from observations of galaxies and quasars, providing complementary insights into isocurvature perturbations beyond those obtained from the CMB angular power spectrum. The figure demonstrates how isocurvature perturbations influence the timing of the transition from a fully neutral IGM to a partially ionized state. Specifically, higher values of $r_\mathrm{CDM}$ result in an earlier onset of reionization because enhanced small-scale perturbations accelerate the formation of the first luminous structures that emit ionizing photons. Consequently, the neutral hydrogen fraction $x_\mathrm{HI}$ decreases more rapidly compared to scenarios with lower $r_\mathrm{CDM}$ values.

Observational constraints indicate that adiabatic perturbations dominate the large-scale density field, but a small fraction of isocurvature perturbations cannot be ruled out. Since these constraints are derived from the post-reionization Universe, using probes of structure formation before reionization—such as the 21cm signal—provides complementary sensitivity to isocurvature modes. This multi-epoch approach strengthens our ability to detect or further constrain isocurvature perturbations.

\section{Cosmological 21cm signal}
\label{Cosmological 21cm signal}
The 21cm line signal emitted by neutral hydrogen serves as a powerful probe not only during the epoch of reionization but also before it. Below, we summarize the fundamentals of the 21cm line signal. The 21cm signal provides a unique and complementary probe of the early universe, particularly during the epochs before reionization and on small spatial scales, where traditional CMB and galaxy observations have limited sensitivity. By exploring a wider parameter space, including larger isocurvature fractions, we can understand how the 21cm signal responds to these initial conditions, clarify the discriminating power of various statistical measures.

\subsection{21cm power spectrum}

The differential brightness temperature for the 21cm line can be expressed as follows \citep{FURLANETTO2006181}:

\begin{eqnarray}
\delta T_b(z) &=& \frac{T_S - T_{\gamma}}{1+z}(1 - \mathrm{e}^{-\tau_{\nu_0}}) \nonumber \\
&\approx& 27x_{\rm HI}(1+\delta_{\rm m}) \left(\frac{H}{dv_r /dr + H}\right) \left(1 - \frac{T_{\gamma}}{T_S}\right) \nonumber \\
&&\times \left(\frac{1+z}{10}\frac{0.15}{\Omega_mh^2}\right)^{\frac{1}{2}} \left(\frac{\Omega_bh^2}{0.023}\right)\ \mathrm{[mK]},
\end{eqnarray}

where $T_S$ is the spin temperature, $T_{\gamma}$ is the temperature of the CMB, $x_{\mathrm{HI}}$ is the fraction of neutral hydrogen, $\delta_m$ is the matter overdensity, $H$ is the Hubble parameter, and $v_r$ is the peculiar velocity along the line of sight. This equation highlights the dependence of the 21cm brightness temperature on various physical properties, such as the state of hydrogen ionization, the thermal history of the IGM, and the large-scale structure of the Universe.

The power spectrum $P(\mathbf{k})$ of the 21cm signal is a crucial statistical tool and is defined as:

\begin{eqnarray}
\langle \widetilde{\delta T_b}(\mathbf{k}_1,z)\,\widetilde{\delta T_b}(\mathbf{k}_2,z)\rangle = (2\pi)^3 \delta_D(\mathbf{k}_1+\mathbf{k}_2)\,P_{21}(\mathbf{k}_1,z),
\end{eqnarray}


where $\langle \cdots \rangle$ denotes the ensemble average, $\delta_D(\mathbf{k})$ is the Dirac delta function, and $\widetilde{\delta T_b}(\mathbf{k},z)$ represents the Fourier transform of the 21cm brightness temperature fluctuation $\delta T_b(\mathbf{x},z)$. The power spectrum captures the spatial correlations of the 21cm signal, offering insights into the distribution of matter and the properties of the IGM during key cosmic epochs.

The 21cm line power spectrum is a powerful probe of the early Universe, enabling detailed studies of the distribution and properties of neutral hydrogen during different epochs, such as the CD and the EoR \citep{2008MNRAS.384.1069B,2009MNRAS.393.1449H,2014MNRAS.443.3090W,2015MNRAS.451..467S}. By analyzing the power spectrum, we can investigate the astrophysical processes governing star formation, X-ray heating, and the ionization of the IGM. Furthermore, the power spectrum is sensitive to the fundamental physics of the early Universe, including dark matter properties and initial conditions for structure formation.

In this study, we utilize the publicly available semi-numerical simulation code \texttt{21cmFAST} \citep{2007ApJ...669..663M,2011MNRAS.411..955M} to simulate the cosmic 21cm line signal. This code efficiently generates large-scale 21cm signal maps, including brightness temperature maps, ionized fraction distributions, and power spectra. Our simulations are performed with a box size of 300 cMpc and $200^3$ pixel grids, achieving a resolution of 1.5 cMpc per pixel. The simulations span redshifts from $z=30$ to $z=6$, capturing the evolution of the 21cm signal over a wide range of epochs and scales. These settings are essential for studying the interplay between small- and large-scale features in the 21cm signal and their connection to cosmic history.

To evaluate the robustness of the constraints on isocurvature fluctuation, we employ three astrophysical models, summarized in Table~\ref{tab:pure_astro_params_3models}. These models represent different parameter sets constrained by HERA observations. The key parameters include the $\alpha_*$ (index of the stellar-to-halo mass relation), $M_\mathrm{turn}$ (minimum halo mass for star formation), $t_*$ (normalized star formation timescale), and the X-ray luminosity-to-star formation rate ratio (a comprehensive explanation of the model \citep{Park_2019}). Each of these parameters significantly influences the 21cm global signal and its power spectrum. In our framework, we do not introduce a single star formation efficiency (SFE) parameter; instead, the effective star-formation efficiency is controlled jointly by $(\alpha_*, M_\mathrm{turn}, t_*)$. Intuitively, for halos above $M_\mathrm{turn}$ the efficiency increases roughly with halo mass as $(M_h/M_\mathrm{turn})^{\alpha_*}$, while $t_*$ sets the overall normalization (shorter $t_*$ implies higher effective SFE at fixed $M_h$). Hence, decreasing $M_\mathrm{turn}$ or $t_*$ raises the population-averaged SFE by activating more low-mass halos or accelerating star formation, whereas increasing $\alpha_*$ tilts star formation toward higher-mass halos. The X-ray luminosity-to-SFR ratio primarily governs the timing and uniformity of IGM heating: larger values drive earlier and more spatially uniform heating (reducing temperature contrast and weakening the variance peak associated with X-ray heating), whereas smaller values produce slower, patchier heating that can enhance that peak. Because higher effective SFE and blue-tilted/isocurvature-enhanced small-scale power both advance key milestones (WF coupling, X-ray heating, reionization), their observable signatures can be partially degenerate; our use of both variance and skewness across redshift helps to disentangle these effects. For reference, among our three models, Model~2 adopts smaller $M_\mathrm{turn}$ and $t_*$ and a larger X-ray luminosity-to-SFR ratio than Model~1 (higher effective SFE and earlier, more uniform heating), while Model~3 adopts a larger $M_\mathrm{turn}$ and a lower X-ray luminosity-to-SFR ratio (lower effective SFE, delayed and patchier heating) \citep{2022PhRvD.105h3523M}. Unless otherwise stated, all figures showing the dependence on the isocurvature parameters adopt astrophysical model 1 in Table~I as the fiducial astrophysical setup.

\begin{table}
\begin{center}
\small
\begin{tabular}{|c|c|c|c|c|} 
 \hline
 & $\alpha_*$ & $M_\mathrm{turn}~[M_\odot]$ & $t_*$ & 
 \makecell{$\log_{10} \left( \frac{L_{\mathrm{X<2.0keV}}}{\mathrm{SFR}} \right)$ \\ $[\mathrm{erg~s}^{-1} M_\odot^{-1}~\mathrm{yr}^{-1}]$} \\ 
 \hline
 model 1 & 0.50 & $3.8 \times 10^8$ & 0.60 & 40.64 \\ 
 \hline
 model 2 & 0.41 & $1.6 \times 10^8$ & 0.29 & 41.52 \\ 
 \hline
 model 3 & 0.62 & $1.5 \times 10^9$ & 0.86 & 39.47 \\ 
 \hline
\end{tabular}
\caption{
Astrophysical parameters for the three models. Model 1 represents the mean values constrained by HERA observations, while models 2 and 3 correspond to the 1$\sigma$ limits.}
\label{tab:pure_astro_params_3models}
\end{center}
\end{table}

The top panel of Fig. \ref{fig:3kindps} compares the three astrophysical models listed in Table \ref{tab:pure_astro_params_3models}. The peaks of the 21cm power spectrum from right to left correspond to different astrophysical effects (Wouthuysen-Field effect, X-ray heating and reionization) \citep{2023PASJ...75S...1S}, and these astrophysical parameters modulate both the timing and amplitude of these peaks. By comparing the outputs of these models, we can explore how deviations from adiabatic conditions manifest in the 21cm signal. This approach provides a framework to isolate isocurvature contributions and refine our understanding of the early Universe.

\subsection{One-point statistics}
The motivation for using one-point statistics is twofold. 
First, they provide simple but physically informative summaries of the 21-cm brightness-temperature distribution across redshift. 
The variance mainly traces the overall fluctuation amplitude and the timing of major transitions, such as Wouthuysen--Field coupling, X-ray heating, and reionization. 
The skewness, by contrast, is sensitive to asymmetry and non-Gaussian structure in the brightness-temperature distribution, and therefore provides complementary information on patchy heating and ionization that is not captured by the variance alone. 
Second, one-point statistics can be measured directly from smoothed image maps and offer an alternative route to parameter inference that complements, rather than duplicates, standard power-spectrum analyses.

The variance of a continuous field can be determined by integrating the power spectrum over all wave numbers. Similarly, the skewness is associated with an integral of the bispectrum over the wave numbers\citep[e.g.][]{2016MNRAS.458.3003S,2016PASJ...68...61K}. These quantities can be expressed mathematically as:

\begin{equation}
\sigma^2 = \int \frac{d^3 k}{(2\pi)^3} P({\bf k}),
\label{eq:variance_continuous}
\end{equation}

\begin{equation}
\gamma = (\overline{\delta T_b})^3 \int \frac{d^3 k_1}{(2 \pi)^3} \int \frac{d^3 k_2}{(2 \pi)^3} B({\bf k_1}, {\bf k_2}, -{\bf k_1} - {\bf k_2}),
\label{eq:skewness_continuous}
\end{equation}

where $P({\bf k})$ is the power spectrum as a function of the wave vector ${\bf k}$, and $B({\bf k_1}, {\bf k_2}, -{\bf k_1} - {\bf k_2})$ is the bispectrum that characterizes the three-point correlations of the field. The integral over wave numbers ensures that the variance $\sigma^2$ and skewness $\gamma$ account for contributions from fluctuations on all spatial scales.

For discrete data, such as the pixelized 21cm brightness temperature maps produced in numerical simulations or observational data, the variance and skewness are computed differently. They are commonly defined as:

\begin{eqnarray}
\sigma^2 &=& \frac{1}{N} \sum_{i=1}^{N} \big[ X_i - \overline{X} \big]^2,
\label{eq:variance_discrete} \\
\gamma &=& \frac{1}{N \sigma^3} \sum_{i=1}^{N} \big[ X_i - \overline{X} \big]^3,
\label{eq:skewness_discrete}
\end{eqnarray}
where $X_i$ is the value of the variable (e.g., the 21cm brightness temperature) in the $i$-th pixel, $\overline{X}$ is the mean value of $X$, and $N$ is the total number of pixels in the map.

The skewness $\gamma$ provides a measure of asymmetry in the distribution of values. A negative skewness indicates that the distribution has a tail extending towards lower values, while a positive skewness implies a tail extending towards higher values. Variance and skewness are key statistical descriptors that encapsulate different aspects of the underlying distribution. In the context of 21cm cosmology, these one-point statistics are particularly useful for probing the overall amplitude of fluctuations and non-Gaussian features of the signal. In this work, we focus on the redshift evolution of these two one-point
statistics as our forecasting observables. This allows us to quantify the
standalone constraining power of the low-order moments of the 21-cm
brightness-temperature field. Although more information-rich observables,
such as the full 21-cm power spectrum or bispectrum, may further improve
the constraints, they require additional modeling of covariance,
foregrounds, and survey-dependent systematics. We therefore leave a joint
analysis including these observables to future work.

\subsection{Thermal noise}

Observational errors, particularly instrumental noise, play a critical role in determining the sensitivity of variance and skewness measurements to underlying physical parameters. While foreground noise which is beyond the scope of this paper is neglected in this analysis for simplicity, instrumental noise is explicitly considered to ensure the reliability of parameter constraints derived from the 21cm signal. The instrumental noise on the brightness temperature, $\Delta T^N$, measured by an interferometer is given by \citep{FURLANETTO2006181}:

\begin{equation}
\Delta T^N = \frac{T_{\rm sys}}{\eta_{\rm f}\sqrt{\Delta\nu t_{\rm int}}},
\label{eq:InstNoise1}
\end{equation}
where $T_{\rm sys}$ is the system temperature, primarily determined by the sky temperature in the radio-quiet regions of the sky. It follows the relation $T_{\rm sys}=180\,(\nu/180\ {\rm MHz})^{-2.6}\,{\rm K}$ \citep{1982A&AS...47....1H}. The array filling factor, $\eta_{\rm f}$, is defined as $\eta_{\rm f}=A_{\rm tot}/D^2_{\rm max}$, where $A_{\rm tot}$ is the total effective area of the array and $D_{\rm max}$ is the maximum baseline.

The brightness temperature noise, $\sigma_{\rm noise}$, is expressed as \citep{2014MNRAS.443.3090W}:

\begin{eqnarray}
\sigma_{\rm noise}
&=& 0.37\ {\rm mK} \left(\frac{10^6\ {\rm m}^2}{A_{{\rm tot}}}\right)
                  \left(\frac{5^{'}}{\Delta \theta}\right)^{2}
                  \left(\frac{1+z}{10}\right)^{4.6} \nonumber \\
& & \times \sqrt{\left(\frac{1~{\rm MHz}}{\Delta \nu}
                       \frac{1000~{\rm hours}}{t_{\rm int}}\right)},
\label{eq:noise}
\end{eqnarray}
where $\Delta\theta$ is the angular resolution of the interferometer, $\Delta \nu$ is the frequency resolution, and $t_{\rm int}$ is the total observation time.

Instrumental noise, characterized by $\sigma_{\rm noise}$, determines the precision of variance and skewness measurements. The parameters influencing $\sigma_{\rm noise}$ include:
- $A_{\rm tot}$: Larger total effective area reduces noise and enhances sensitivity.
- $\Delta\theta$: Finer angular resolution enables us to explore the spatial distribution of the brightness temperature with higher precision but increases noise due to smaller beam size.
- $\Delta\nu$: Higher frequency resolution allows for finer spectral features.
- $t_{\rm int}$: Longer integration times reduce noise as $\propto t_{\rm int}^{-1/2}$.

Modeling instrumental noise accurately ensures that parameter constraints reflect the true detectability of the 21cm signal rather than being dominated by observational artifacts. This consideration is essential for interpreting variance and skewness in terms of the physical processes driving the evolution of the early Universe.

\subsection{Fisher forecast}

The Fisher matrix plays a crucial role in parameter estimation by quantifying the curvature of the likelihood surface around the maximum likelihood point. Our Fisher analysis is intentionally restricted to one-point statistics, namely the variance and skewness as functions of redshift. 
We therefore do not include the full 21-cm power spectrum in the Fisher matrix in this work. 
The reason is that the aim of this paper is to quantify the constraining power of one-point statistics alone.  The components of the Fisher matrix are defined as:

\begin{equation}
\mathcal{F}_{ij} = - \left\langle \frac{\partial^2 \ln{\mathcal{L}}}{\partial \theta_i \partial \theta_j} \right\rangle,
\end{equation}

where \( \mathcal{L}(\mathbf{\theta}) \) represents the likelihood function of the model parameters \( \mathbf{\theta} \). According to the Cramér-Rao theorem, the inverse of the Fisher matrix sets a lower bound on the covariance of any unbiased estimator of \( \mathbf{\theta} \). This inverse, therefore, establishes a theoretical limit on the precision with which model parameters can be estimated from future observational data \citep{2009arXiv0906.4123C,Verde2010}.

For practical implementation, we calculate the Fisher matrix elements as follows:

\begin{equation}
\mathcal{F}_{ij}
= \sum_{k=1}^N \frac{1}{\sigma_k^2}
               \frac{\partial x_k(\vec{p})}{\partial p_i}
               \frac{\partial x_k(\vec{p})}{\partial p_j}
               \Bigg|_{\vec{p}=\vec{p}_{\rm fid}},
\end{equation}

where \( x_k(\vec{p}) \) is the observable quantity dependent on the model parameters \( \vec{p} \) and $k$ denotes the redshift bin. This formulation assumes that the likelihood function follows a Gaussian distribution and that the data points are statistically independent. In this work, we adopt uncorrelated errors for analytical simplicity. In our analysis, we adopt the variance and skewness as \( x_k(\vec{p}) \). \( \sigma_k \) represents the corresponding observational uncertainty. The summation is performed over all independent data points.

The inverse of the Fisher matrix, denoted as \( \mathcal{C} = \mathcal{F}^{-1} \), provides the covariance matrix of the parameter estimates. Consequently, the forecasted uncertainty for the \( i \)-th parameter is given by:

\begin{equation}
\sigma(\theta_i) = \sqrt{\mathcal{C}_{ii}}.
\end{equation}

These uncertainties are valid in the vicinity of the fiducial model and assume that the model accurately describes the data.

In our Fisher analysis, we adopt fiducial parameter values of
$r_{\rm CDM}=0.05$, $n_{\rm iso}=2.5$, together with astrophysical model~1 in Table~I, $\alpha_\star=0.50$,
$M_{\rm turn}=3.8\times10^8\,M_\odot$, $t_\star=0.60$, and
\[
\ell_X \equiv
\log_{10}\!\left(
\frac{L_{X<2.0\,{\rm keV}}/{\rm SFR}}
{{\rm erg}\ {\rm s}^{-1}\ M_\odot^{-1}\ {\rm yr}}
\right)
=40.64.
\]
The isocurvature fiducial is chosen to lie below the conservative CMB-motivated upper envelope $r_{\max}(n_{\rm iso})$ discussed above, while still producing a clearly visible response in the 21cm observables. This makes it suitable for illustrating the sensitivity of variance and skewness to isocurvature perturbations without overstating the precision of the currently available CMB constraints.

To ensure the convergence of the numerical derivatives entering the Fisher matrix, we test finite perturbations around the fiducial model. Specifically, we vary $r_{\rm CDM}$, $\alpha_\star$, $M_{\rm turn}$, and $t_\star$ by $\pm 3\%$, $n_{\rm iso}$ by $\pm 0.5\%$, and
$\ell_X$ 
by $\pm 0.1\%$. These step sizes are chosen to balance numerical stability and local linearity, thereby allowing us to derive robust forecasted constraints around the fiducial model.

\section{Results}
\label{Sec3}
\subsection{Power spectrum}
\begin{figure}
    \includegraphics[width=1.0\hsize]{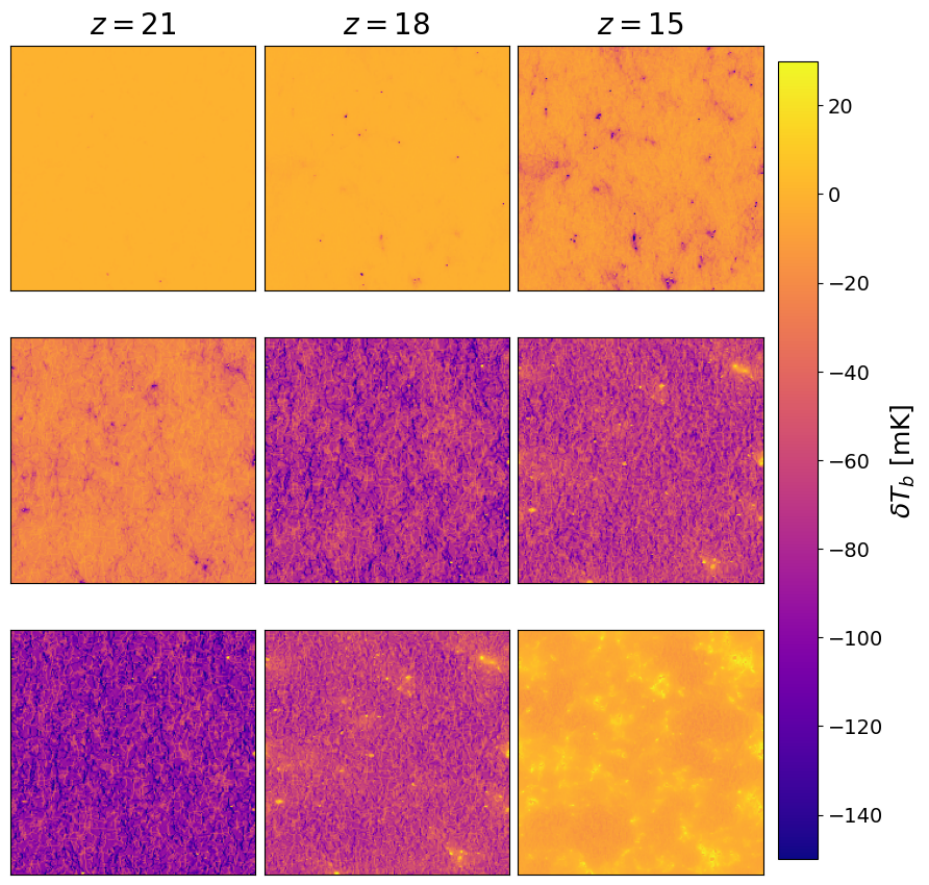}
    \caption{ From top to bottom: maps of $\delta T_b$ for $r_{\rm CDM}=0$, $0.05$, and $0.1$, respectively. Here we fix $n_{\rm iso}=2.5$ and adopt astrophysical model~1 in Table~I. From left to right: $z=21$, $18$, and $15$.}
  \label{fig:3kindmapsoftb}
\end{figure}

In Fig. \ref{fig:3kindmapsoftb}, we first show the maps of the 21cm brightness temperature $\delta T_b$ for different values of the isocurvature perturbation ratio $r_{\text{CDM}}=0, 0.05, 0.1$ at three different redshifts: $z=21$, $z=18$, and $z=15$, with a fixed value of the spectral index $n^{\text{iso}}=2.5$. Warmer (more orange) regions in the maps indicate higher \( \delta T_b \) values, while cooler (purple) regions correspond to lower \( \delta T_b \). At \( r_{\mathrm{CDM}} = 0 \) (top row), the distribution is dominated purely by adiabatic fluctuations, resulting in relatively smoother structures. As \( r_{\mathrm{CDM}} \) increases to 0.05 (middle row) and 0.1 (bottom row), increasingly pronounced small-scale fluctuations in \( \delta T_b \) emerge, particularly at the lower redshifts (rightmost panels).

When an isocurvature component is added alongside the usual adiabatic fluctuations, isocurvature perturbations accelerate structure formation in the universe. This is because they promote faster growth of density contrasts, leading to earlier collapse of matter into structures like galaxies and halos. As a result, the spatial inhomogeneities introduced by isocurvature modes become more pronounced, particularly at lower redshifts where gravitational clustering and non-linear growth processes are more efficient. This accelerated structure formation directly contributes to the enhanced small-scale features observed in the 21 cm brightness temperature maps, providing a clear signature of the isocurvature component in the primordial fluctuations.

Furthermore, isocurvature perturbations primarily boost the matter density contrast on small scales in our simulations. Since the 21cm signal depends sensitively on the underlying gas density and temperature, this enhanced small-scale clustering manifests as stronger contrast in \( \delta T_b \) maps (Fig.~\ref{fig:3kindmapsoftb}), particularly for larger values of \( r_{\mathrm{CDM}} \). Regions with higher CDM overdensities can influence the surrounding gas by altering its gravitational potential and thermal evolution. As a result, the 21cm brightness temperature maps show patches of enhanced or diminished intensity, with enhancement being the dominant trend in our results. 

Overall, adding even a modest fraction of isocurvature perturbations (\( r_{\mathrm{CDM}} \neq 0 \)) increases the spatial inhomogeneity of hydrogen gas density and temperature, producing more pronounced small-scale structure in the \( \delta T_b \) maps compared to the purely adiabatic case. These differences become increasingly evident at lower redshifts (e.g., \( z = 15 \)) as non-linear growth further amplifies the initial perturbations.

\begin{figure}[t]
    \centering
   \includegraphics[width=0.85\hsize]
{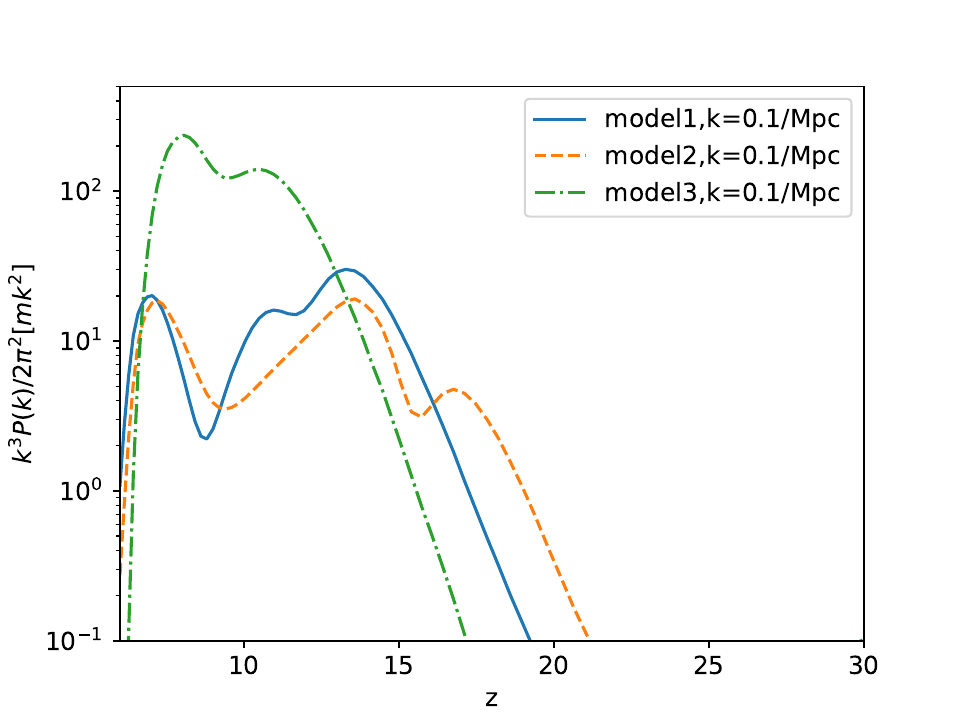}
\includegraphics[width=0.85\hsize]
{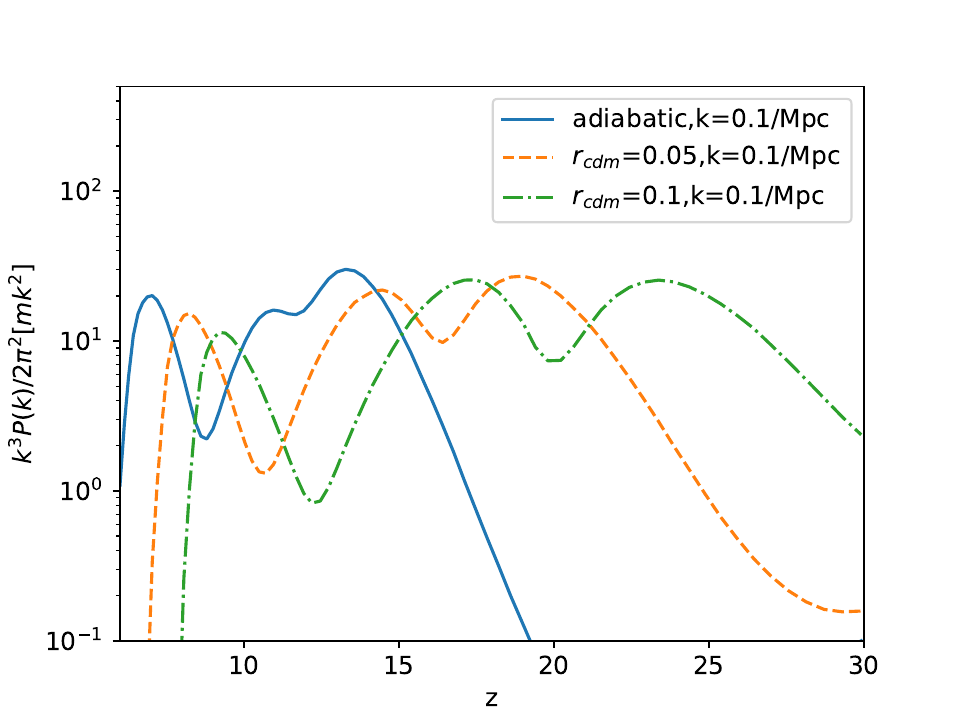}
\includegraphics[width=0.85\hsize]{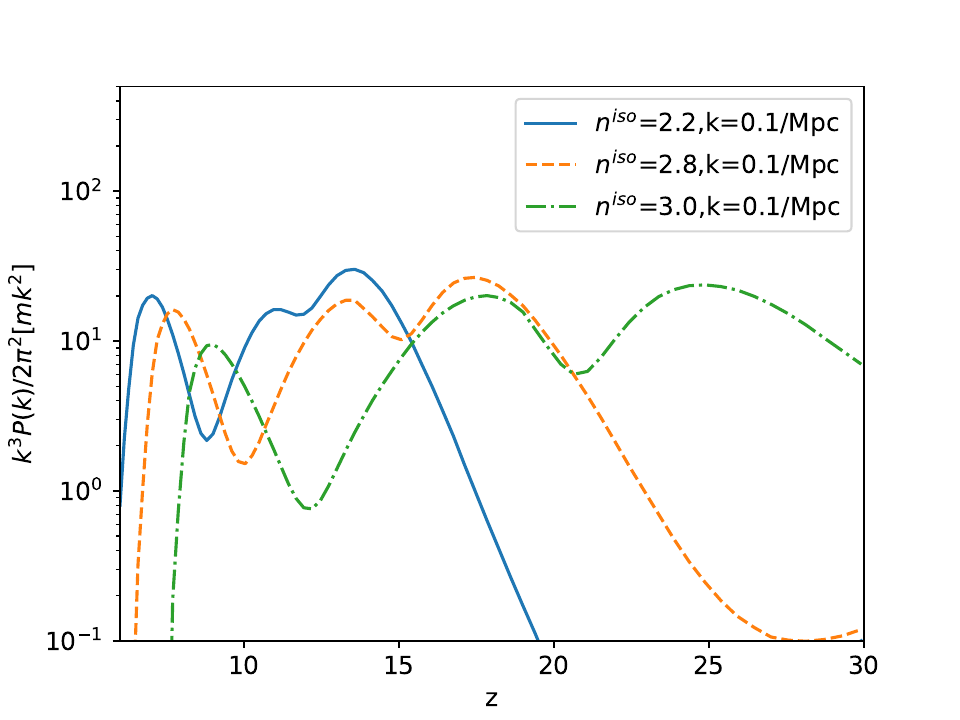}
    \caption{ Illustration of the 21cm power spectrum.
    Top: comparison of the three astrophysical models listed in Table~I.
    Middle: varying $r_{\rm CDM}=0$, $0.05$, and $0.10$ at fixed $n_{\rm iso}=2.5$ and fixed astrophysical model~1.
    Bottom: varying $n_{\rm iso}=2.2$, $2.8$, and $3.0$ at fixed $r_{\rm CDM}=0.05$ and fixed astrophysical model~1.
    All curves are shown at $k=0.1\,{\rm Mpc}^{-1}$. }
    \label{fig:3kindps}
\end{figure}

To analyze the 21cm brightness temperature image map, we first calculate the 21cm power spectrum.  Fig.~\ref{fig:3kindps} demonstrates how variations in astrophysical and isocurvature parameters affect the 21cm power spectrum. The top panel compares three astrophysical scenarios (Table~\ref{tab:pure_astro_params_3models}) differing mainly in star formation efficiency and X-ray heating efficiency. These changes produce substantial variations in both the amplitude and shape of the power spectrum, shifting the timing and intensity of its key peaks. This highlights the strong influence of stellar and X-ray heating processes on the evolution of the 21cm signal.

The middle and bottom panels explore the effects of isocurvature perturbations by varying the amplitude $r_\mathrm{CDM}$ and the spectral index $n^\mathrm{iso}$. Each peak in these panels corresponds to a specific astrophysical process, such as Wouthuysen–Field coupling, X-ray heating, or reionization \citep{2023PASJ...75S...1S}. Introducing isocurvature fluctuations enhances the formation of small-scale structures, triggering these processes earlier and shifting all characteristic peaks to higher redshifts.

While increasing $r_\mathrm{CDM}$ or $n^\mathrm{iso}$ shifts the peaks markedly, it leaves the overall shape and amplitude of the power spectrum largely unchanged. This indicates that isocurvature perturbations mainly alter the timing of structure formation rather than the fundamental shape of the 21cm signal. In contrast, astrophysical parameters such as star formation and X-ray heating efficiencies affect both the amplitude and the evolutionary pattern. The 21cm power spectrum during the Cosmic Dawn and EoR is therefore shaped by the interplay between cosmological initial conditions and astrophysical processes.

\begin{figure}
    \includegraphics[width=0.82\hsize]{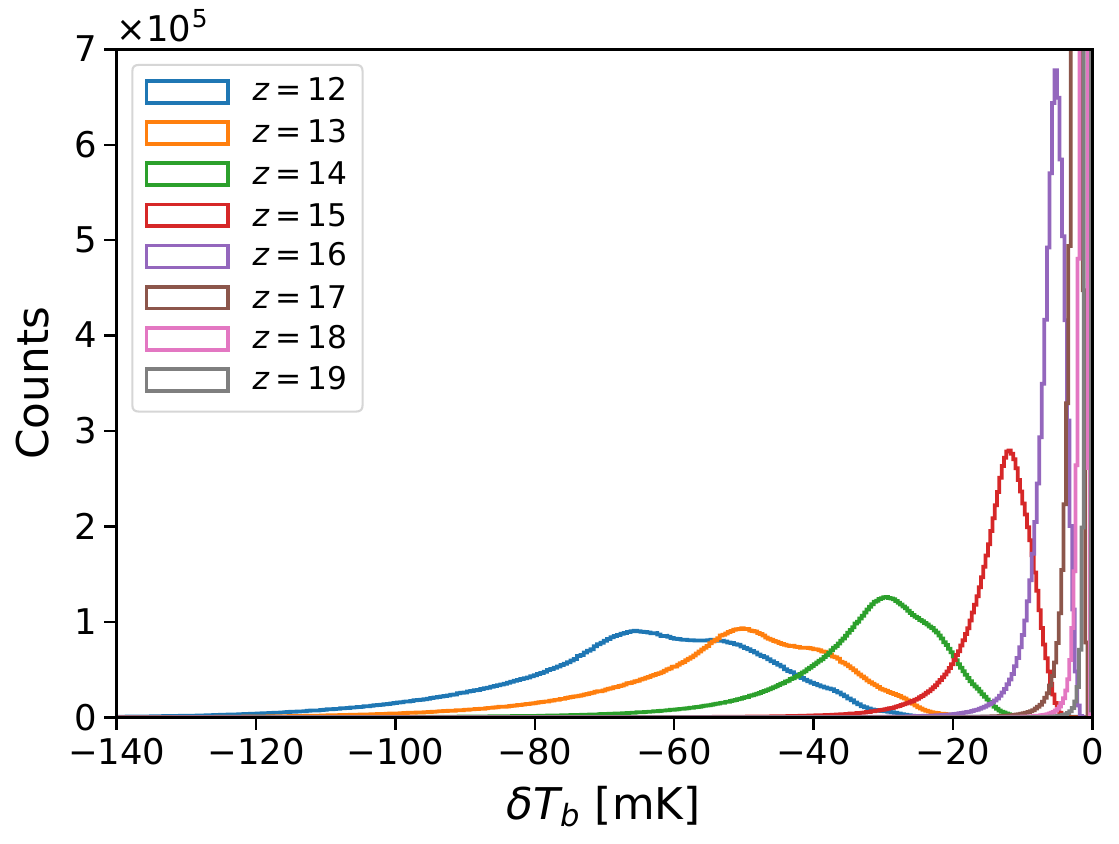}
    \includegraphics[width=0.85\hsize]{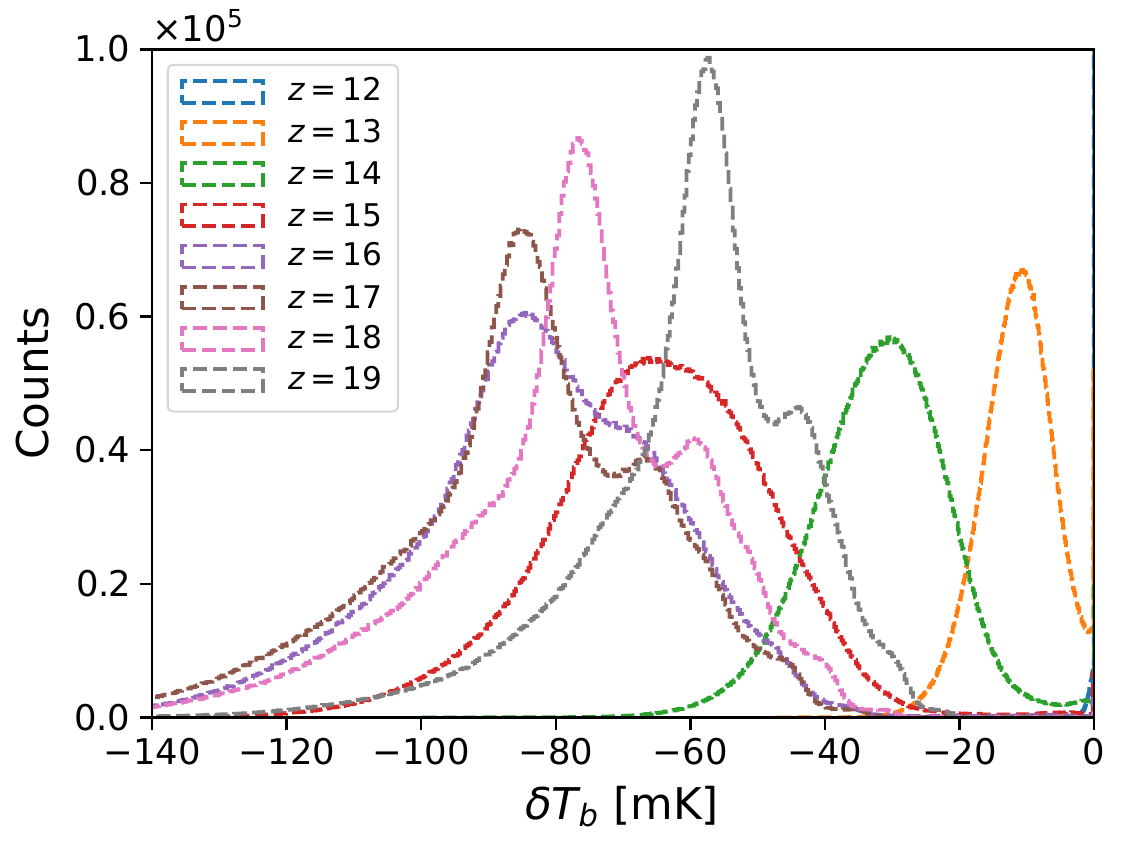}
    \caption{Top: PDF of $\delta T_b$ in the adiabatic case, adopting astrophysical model~1 in Table~I.
    Bottom: PDF of $\delta T_b$ for the isocurvature case with $r_{\rm CDM}=0.05$ and $n_{\rm iso}=2.5$, again adopting astrophysical model~1. }
  \label{fig:3kindpdf}
\end{figure}

\subsection{One-point statistics}

We compare the probability distribution function (PDF) of $\delta T_b$ with and without isocurvature perturbations in Fig.~\ref{fig:3kindpdf}. For nonzero $r_\mathrm{CDM}$, the PDF develops a secondary peak at higher brightness temperatures shortly after the Wouthuysen–Field (WF) effect turns on, typically at redshifts $z \sim 16$–18 in our fiducial models. This feature arises because enhanced small-scale structure, induced by isocurvature perturbations, leads to the early formation of X-ray sources. These sources locally heat the gas, increasing the spin temperature $T_s$ (coupled to the kinetic temperature $T_K$) in those regions and producing higher $\delta T_b$, while less-affected regions remain cooler. The coexistence of these hot and cold regions produces a bimodal temperature distribution: a primary peak from the bulk of cooler regions and a secondary peak from localized, X-ray–heated regions. As cosmic time progresses to lower redshifts ($z \lesssim 14$), X-ray heating becomes more widespread and uniform, reducing the temperature differences between regions. Consequently, the secondary peak diminishes and eventually disappears, while the primary peak shifts to higher $\delta T_b$ values due to the overall rise in $T_s$.

To analyze the 21cm image map more quantitatively, we calculate the variance and skewness of the 21cm image map.

\begin{figure}
    \includegraphics[width=0.85\hsize]{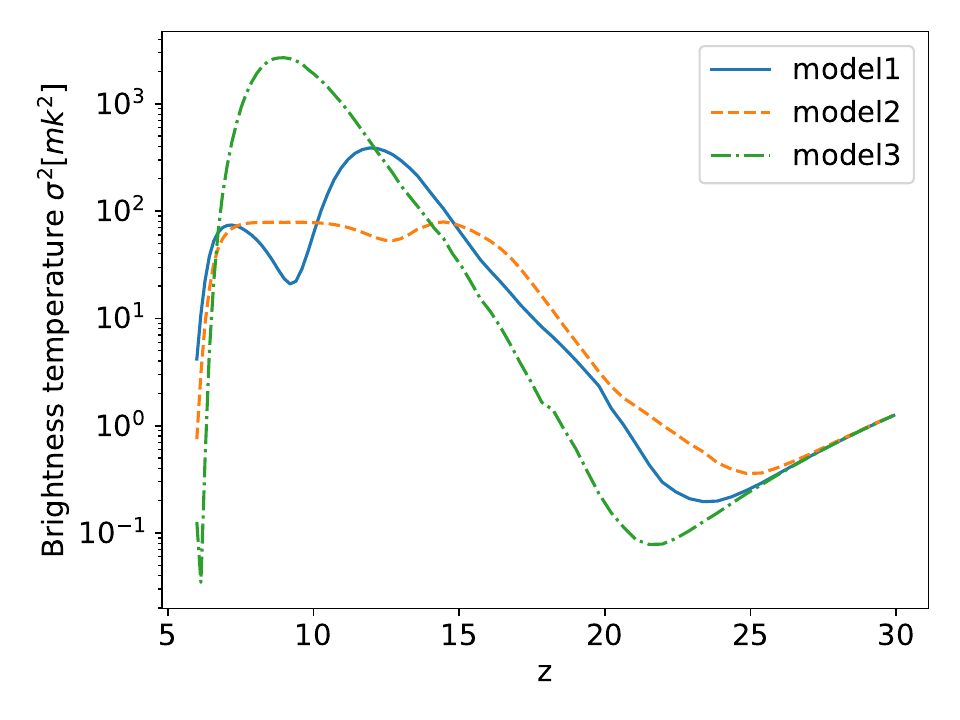}
     \includegraphics[width=0.85\hsize]{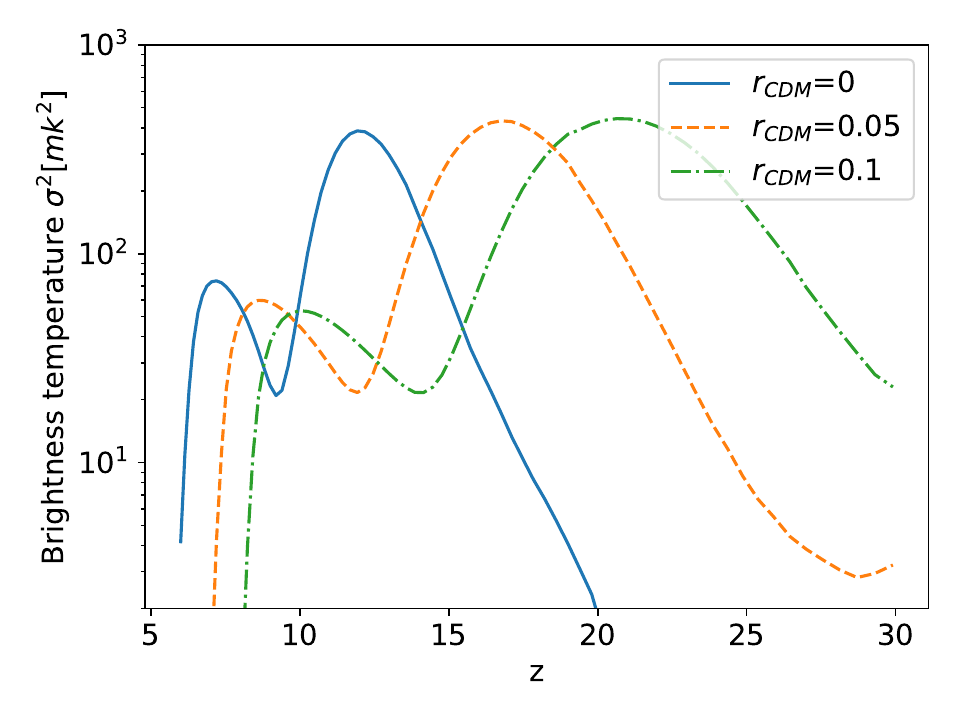}
    \caption{Top: variance of $\delta T_b$ for the three astrophysical models in Table~I. Bottom: variance of $\delta T_b$ for $r_{\rm CDM}=0$, $0.05$, and $0.1$, with $n_{\rm iso}=2.5$ and astrophysical model~1 fixed.}
  \label{fig:var_3all}
\end{figure}

Fig. \ref{fig:var_3all} shows the evolution of the variance of the 21cm brightness temperature, $\delta T_b$, as a function of redshift. The variance, which quantifies the overall amplitude of fluctuations in $\delta T_b$, typically displays two distinct peaks in model 1. The first peak appeared at lower redshift and is associated with the rapid decline in the neutral hydrogen fraction as reionization commences, while the second peak emerges when localized regions begin to experience X-ray heating due to the formation of small-scale structures.

In model~2 (relative to model~1), a smaller $M_{\rm turn}$ and a shorter $t_\ast$ raise the \emph{effective} star-formation efficiency in low-mass halos, and the X-ray luminosity-to-SFR ratio is higher. Although $\alpha_\star$ is lower, this does not necessarily suppress
the low-mass contribution, since for halos below the normalization mass a
smaller $\alpha_\star$ gives a shallower mass dependence of $f_\star(M_h)$.
The net effect is therefore earlier source formation and more efficient early
radiative feedback, accelerating the evolution of the neutral hydrogen field
during reionization. Consequently, the reionization-related peak in the variance shifts to higher redshift and becomes less pronounced, because the many low-mass sources smooth the ionization field.  The larger X-ray luminosity-to-SFR ratio also drives earlier IGM heating, shifting the X-ray–heating peak in the variance to higher redshift. At the same time, the increased uniformity of heating reduces the temperature contrast between hot and cold regions, which can lower the peak amplitude and smooth the redshift evolution of the variance at later times.

In Model~3, the X-ray luminosity-to-SFR ratio is lower than in the other models. This implies that, for a given star formation rate, fewer X-ray photons are produced, delaying and reducing the overall heating of the IGM. However, because heating proceeds slowly and non-uniformly, large cold regions coexist with localized hot regions for an extended period. This strong temperature contrast produces a prominent X-ray heating peak in the variance of the 21cm brightness temperature. 

Furthermore, in Model~3, the minimum halo mass for star formation ($M_{\rm turn}$) is significantly larger than in the other models. A higher $M_{\rm turn}$ confines star formation to more massive halos, effectively reducing the contribution from low-mass halos to the ionizing photon budget. As a result, the production of ionizing photons during the early stages of reionization is suppressed, delaying the reionization-related variance peak to much lower redshifts. In fact, this peak is shifted outside the redshift range shown here, and therefore does not appear in our plots. At the same time, the restriction of star formation to rare, massive halos leads to a more biased and patchy distribution of ionizing sources, which enhances the variance associated with reionization compared to the other models.

\begin{figure}
    \includegraphics[width=0.85\hsize]{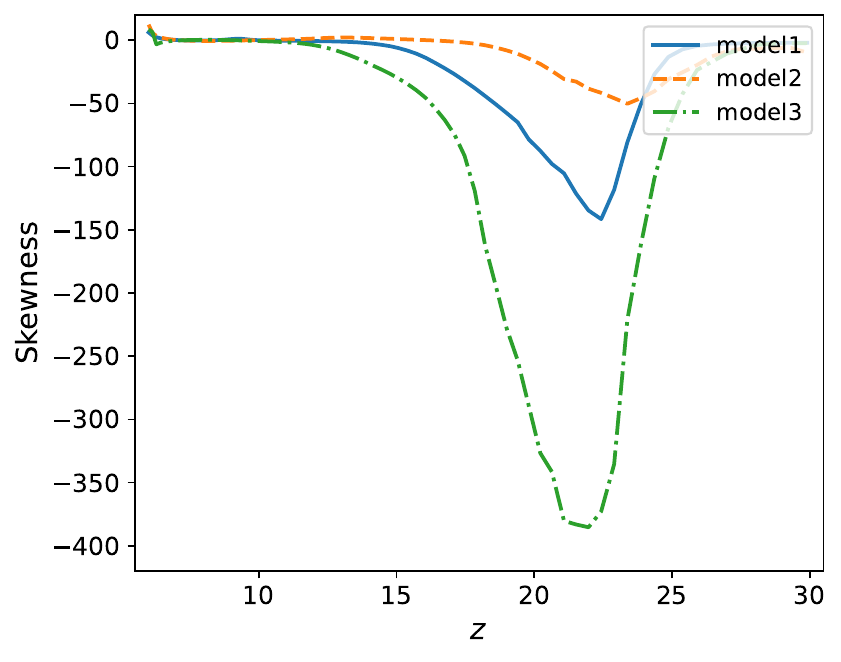}
    \includegraphics[width=0.85\hsize]{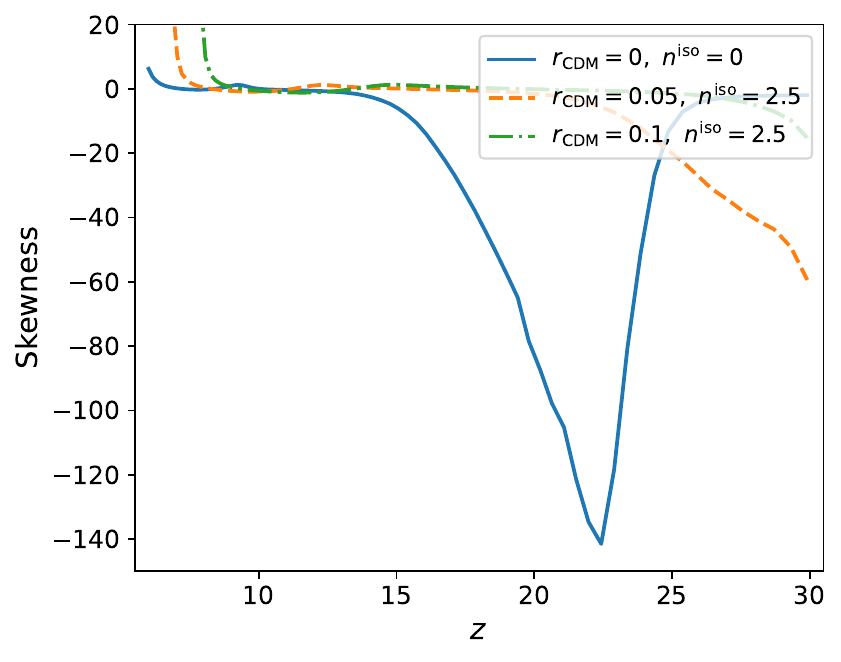}
    \caption{Top: skewness of $\delta T_b$ for the three astrophysical models in Table~I. Bottom: skewness of $\delta T_b$ for $r_{\rm CDM}=0$, $0.05$, and $0.1$, with $n_{\rm iso}=2.5$ and astrophysical model~1 fixed.}
  \label{fig:skew_3all}
\end{figure}

We next see the impacts of isocurvature perturbations on the variance in the bottom panel of Fig.~\ref{fig:var_3all}. The bottom panel of Fig.~\ref{fig:var_3all} illustrates how varying the isocurvature perturbation fraction ($r_{\mathrm{CDM}}$) affects the variance of the 21cm brightness temperature. Increasing $r_{\mathrm{CDM}}$ systematically shifts the peaks associated with reionization and X-ray heating to higher redshifts, indicating that these events occur earlier. For example, increasing \( r_{\mathrm{CDM}} \) from 0 to 0.1 shifts the reionization-related variance peak by $\Delta z \approx 2.8$ and the X-ray heating peak by $\Delta z \approx 8.7$. This trend arises because isocurvature perturbations enhance small-scale density fluctuations, accelerating halo formation, triggering earlier star formation, and thus advancing ionization and heating. The amplitudes of these peaks remain nearly unchanged, showing that isocurvature perturbations primarily affect the timing of these events rather than their strength or detailed shape.

In contrast, changing astrophysical parameters (top panel of Fig.~\ref{fig:var_3all})—such as star formation efficiency, the minimum halo mass for star formation ($M_{\mathrm{turn}}$), or the X-ray luminosity-to-SFR ratio—affects not only the redshift position of the peaks but also their amplitudes and overall shapes. While both cosmological and astrophysical parameters influence the timing of the variance peaks, astrophysical parameters also modify their amplitude and shape, clearly distinguishing their impact from that of isocurvature perturbations. Nonetheless, both \(r_{\mathrm{CDM}}\) and X-ray heating efficiency can shift the variance peaks in similar ways, leading to partial degeneracy that requires joint analysis to resolve.

Figure~\ref{fig:skew_3all} shows the evolution of the skewness of $\delta T_b$ as a function of redshift. Positive skewness indicates a distribution skewed toward higher temperatures, whereas negative skewness reflects a distribution skewed toward lower temperatures. At high redshift {($z \gtrsim 22$)}, when the IGM is cold and mostly neutral, the distribution is skewed toward lower $\delta T_b$, producing negative skewness. As the Universe evolves through the WF coupling and X-ray heating phases, localized heating—particularly from early X-ray sources—introduces a high-temperature tail, driving the skewness from negative to positive. The peak in skewness typically occurs near the onset of widespread X-ray heating.

The skewness is highly sensitive to astrophysical heating. Models with stronger X-ray heating (Model 1 and 2) produce larger local temperature enhancements earlier, driving an earlier zero-crossing and rising to positive skewness. Conversely, lower X-ray efficiencies yield a slower evolution and stronger skewness signatures. Increasing $r_{\mathrm{CDM}}$ shifts the skewness peak toward higher redshifts, again reflecting earlier structure formation and heating. However, because skewness is strongly influenced by localized, non-Gaussian features from astrophysical processes, it is difficult to isolate isocurvature effects from astrophysical uncertainties using skewness alone.

To directly examine the interplay between astrophysical assumptions and isocurvature perturbations, we further compare the three astrophysical models under different values of $r_{\rm CDM}$ in Appendix Fig.~\ref{fig:app_interplay_astro_rcdm}. This figure combines the redshift evolution of the 21cm power spectrum, variance, and skewness for the three astrophysical models listed in Table~I, fixing $n_{\rm iso}=2.5$.

We find that the qualitative effect of isocurvature perturbations is robust across the three astrophysical models. Increasing $r_{\rm CDM}$ generally shifts the characteristic features of all three observables toward higher redshift, indicating earlier structure formation, heating, and reionization. However, the detailed amplitudes, peak widths, and skewness evolution remain model dependent. This demonstrates that the combined effect of astrophysical parameters and isocurvature perturbations is not a simple additive offset; rather, the isocurvature component mainly changes the timing of the main features, while the astrophysical model controls their detailed amplitudes and shapes.

Taken together, the variance primarily traces the global timing of key thermal and ionization milestones, whereas the skewness is more sensitive to localized, non-Gaussian heating features. Combining both statistics can help break degeneracies between cosmological parameters such as \(r_{\mathrm{CDM}}\) and astrophysical heating efficiencies. The magnitude of the variance and skewness shifts shown here is large enough to be potentially detectable with SKA Phase 1 sensitivity, provided that foregrounds and systematics can be mitigated.

\subsection{Realistic observational situation}

\begin{figure}
    \includegraphics[width=1.0\hsize]{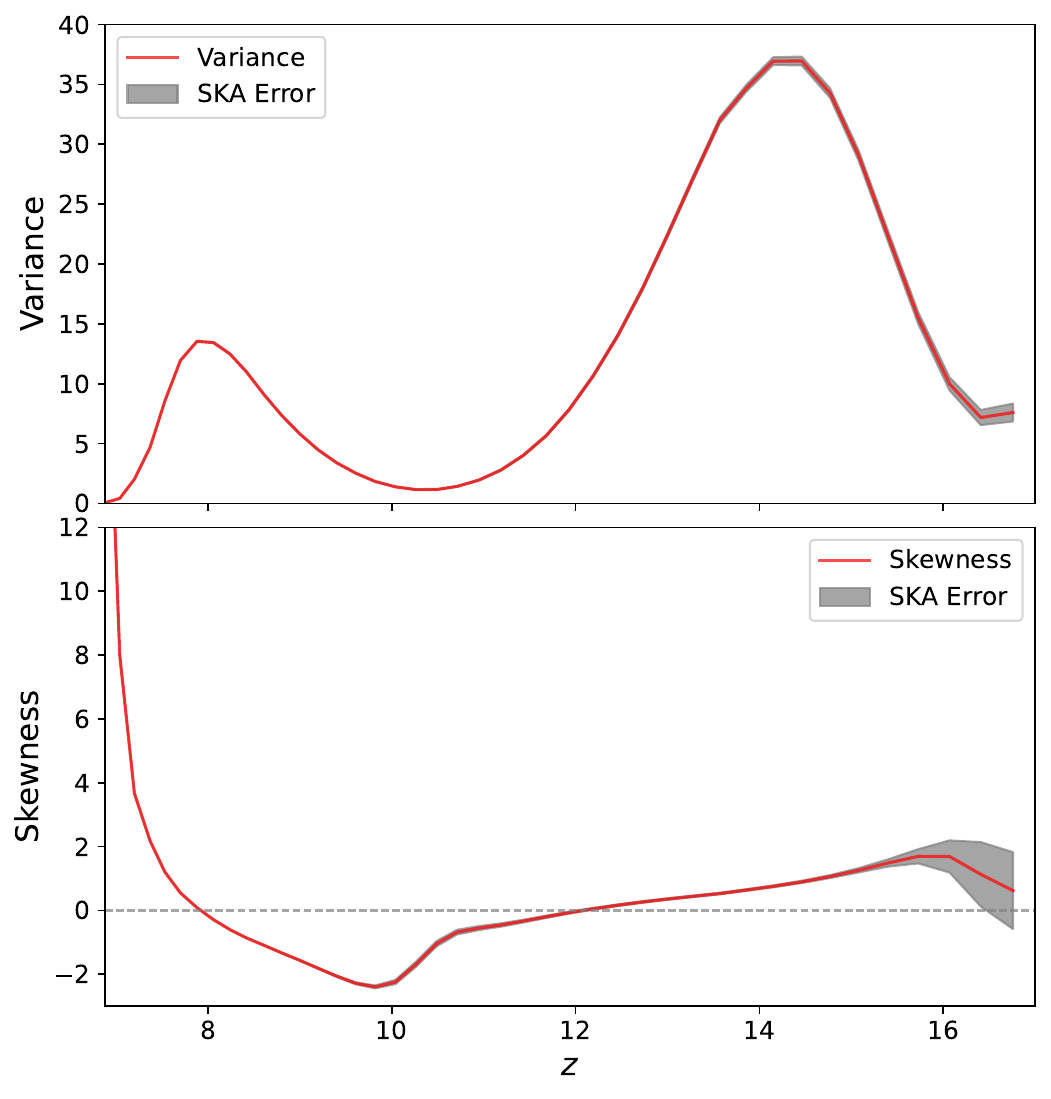}
    \caption{The variance (top) and skewness (bottom) of the brightness temperature for the isocurvature case with $r_{\rm CDM}=0.05$ and $n_{\rm iso}=2.5$, adopting astrophysical model~1 in Table~I, compared with the $1\sigma$ instrumental noise assuming SKA level (shaded region). Both statistics are calculated from smoothed image maps with $R_{\rm smooth}=12\,{\rm Mpc}$ along the redshift direction. This smoothing scale corresponds to the SKA level.} 
  \label{fig:instrument_noise}
\end{figure}

We next consider a more realistic observational situation. At the native 1.5 Mpc resolution of our simulations, pixel-level instrumental noise dominates and completely swamps higher-order statistics such as skewness. To mitigate this, we smooth our 21cm maps to a 12 Mpc scale, roughly matching the SKA beam. This averaging reduces the noise floor by combining many noisy pixels, restoring sensitivity to the cosmic signal on the scales where the array is most effective. Tests with alternative smoothing scales confirm that the qualitative behavior of the variance and skewness evolution is robust, although smaller smoothing scales retain more small-scale information at the expense of higher noise. To explicitly assess the impact of smoothing, Appendix Fig. \ref{fig:appendix_smoothing_scan} shows the redshift evolution of the variance and skewness for $R_{\rm smooth}$=3, 6, 12, and 15 Mpc. We find that the qualitative evolution of both statistics is robust to smoothing. Increasing the smoothing scale mainly reduces their amplitudes, while the overall redshift trends and the locations of the main features remain broadly unchanged. We also include SKA-level thermal noise in the smoothed maps and compute the resulting variance and skewness, as shown in Fig.~\ref{fig:instrument_noise}.

Even after smoothing, the skewness uncertainty exhibits a pronounced bump around \( z \sim 10 \), in contrast to the smoothly varying errors reported by \citet{2014MNRAS.443.3090W}. This feature can be understood from the error propagation of the skewness estimator,
\begin{equation}
    \gamma_3' \;=\;\frac{\hat S_3}{(\hat S_2)^{3/2}},
\end{equation}
whose variance propagates as
\begin{align}
V_{\gamma_3'} &\approx \frac{1}{S_2^3}\,V_{\hat S_3}
+ \frac{9}{4}\,\frac{S_3^2}{S_2^5}\,V_{\hat S_2}
- 3\,\frac{S_3}{S_2^4}\,C_{\hat S_2\hat S_3}.
\end{align}
In our models, the first term, \(V_{\hat S_3}/S_2^3\), dominates. Around \( z \sim 10 \), the second moment \( S_2 \) dips while the variance of the third moment \( V_{\hat S_3} \) rises, producing a local maximum in \( V_{\gamma_3'} \). If \( S_2 \) evolved monotonically, this term would remain smooth and the bump would not appear.

At redshifts \( z \gtrsim 16 \), the 21cm line is observed at very low radio frequencies (below \(\sim\)80 MHz), where diffuse Galactic synchrotron emission dominates the sky temperature. This dramatically increases the system temperature (see Eq.~\ref{eq:InstNoise1}), degrading instrumental sensitivity even for the same integration time. As a result, the pixel-level thermal noise rises steeply at high redshift, and the measurement becomes noise-limited for \( z \gtrsim 15 \).

In our fiducial model, SKA Phase 1 could detect variance and skewness measurements for \( 7 \lesssim z \lesssim 15 \) after 1000 hours of integration. Within this range, the cosmic signal dominates over thermal noise after smoothing, whereas at higher redshifts the measurements are noise-limited. Under realistic noise conditions, complementary statistics such as the bispectrum or one-point PDFs may retain sensitivity to non-Gaussian features even when skewness becomes noise-limited.

The detectability at high redshift could be improved by extending the integration time or by combining observations from multiple low-frequency arrays, potentially mitigating the loss of sensitivity at \( z \gtrsim 16 \). These considerations emphasize that while realistic SKA-level noise and beam smoothing modify the redshift evolution of variance and skewness uncertainties, there remains a substantial redshift window where both statistics can provide valuable constraints on isocurvature perturbations and astrophysical heating processes.

\begin{figure*}[t]
    \centering
    \includegraphics[
        width=0.95\textwidth,
        height=0.82\textheight,
        keepaspectratio]{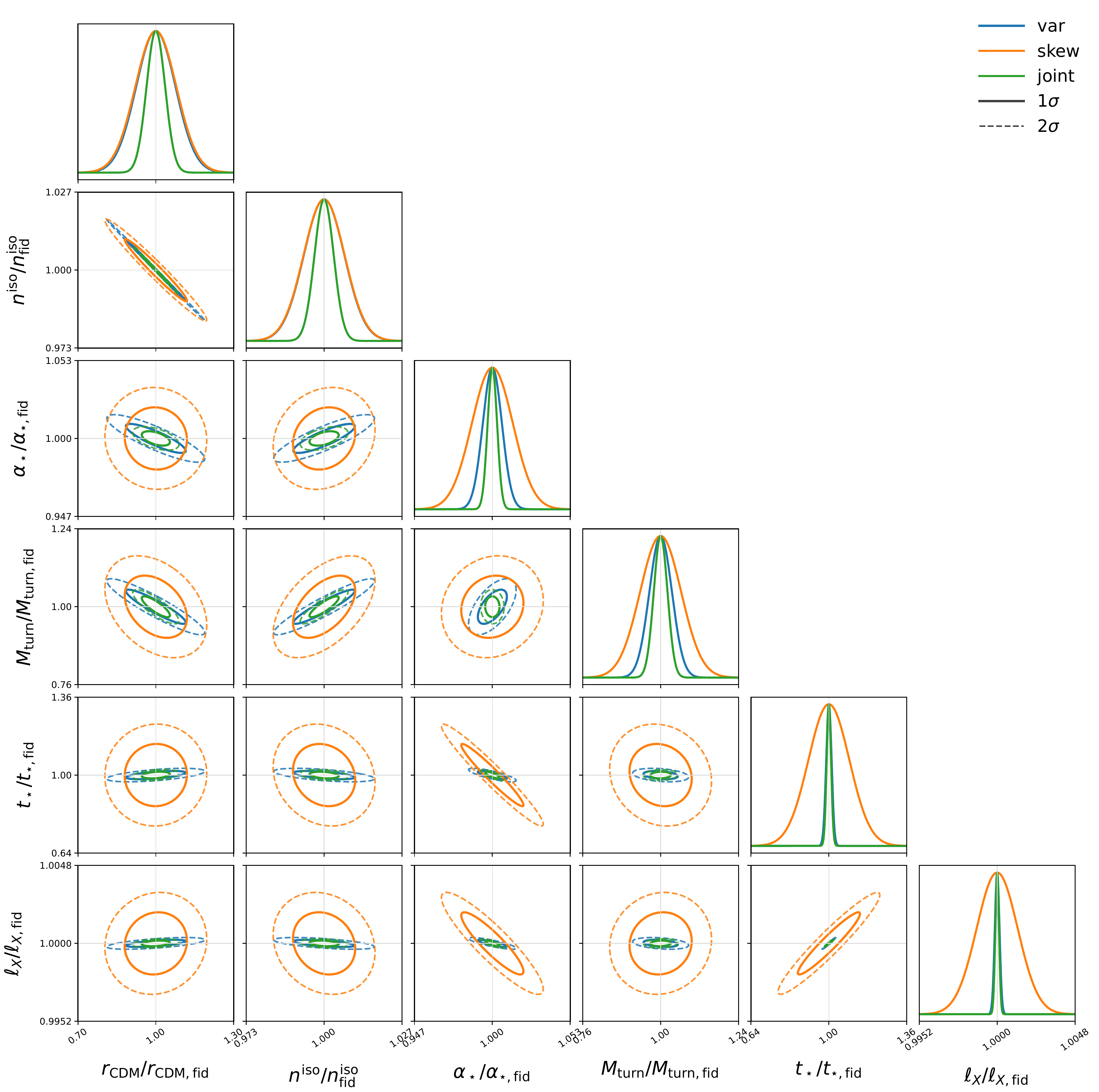}
    \caption{
    Marginalized Fisher constraints on the six parameters.
    The blue, orange, and green contours correspond to the variance-only,
    skewness-only, and joint variance-plus-skewness constraints, respectively.
    Solid and dashed contours denote the $1\sigma$ and $2\sigma$ confidence regions.
    For display purposes, each parameter is normalized by its fiducial value.
    }
    \label{fig:fisher}
\end{figure*}

Figure~\ref{fig:fisher} shows the marginalized Fisher constraints on all six parameters. The one-dimensional constraints are marginalized over the other five parameters, while each two-dimensional contour is marginalized over the remaining four parameters. The blue, orange, and green curves correspond to the variance-only, skewness-only, and joint variance-plus-skewness constraints, respectively. In the off-diagonal panels, solid and dashed contours denote the $1\sigma$ and $2\sigma$ confidence regions. It also makes explicit the dependence on the astrophysical parameters by showing the degeneracies between $(r_{\rm CDM}, n_{\rm iso})$ and $(\alpha_\star, M_{\rm turn}, t_\star, \ell_X)$. 

The figure shows that the joint variance-plus-skewness analysis can constrain the fiducial isocurvature model at the level of
$\Delta r_{\rm CDM}\simeq 1.8\times10^{-3}$ and
$\Delta n_{\rm iso}\simeq 8.4\times10^{-3}$. The variance-only and skewness-only constraints are relatively similar for the two isocurvature parameters, $r_{\rm CDM}$ and $n^{\rm iso}$, but they differ more noticeably for the astrophysical parameters. This discrepancy reflects the differing sensitivities of the two statistics. Variance integrates power over all spatial scales, capturing the bulk amplitude of \(\delta T_b\) fluctuations and thus accumulating high signal-to-noise. Skewness emphasizes localized non-Gaussian tails—such as small, intensely heated regions—making it intrinsically more susceptible to noise. In practice, even after smoothing, the thermal noise increases toward the high-redshift end of the observable range. Since skewness is sensitive to the tails of the brightness-temperature distribution, this increased noise leads to larger skewness uncertainties and weakens the constraining power of skewness-only forecasts. Nonetheless, skewness retains value as a complementary probe: it is particularly sensitive to non-Gaussian features that variance cannot isolate, and it can cross-check or refine variance-based parameter estimates when used in combination.

Both statistics reveal a pronounced degeneracy between $r_{\rm CDM}$ and $n_{\rm iso}$. 
The two parameters do not play the same physical role: $r_{\rm CDM}$ controls the overall amplitude of the isocurvature contribution, whereas $n_{\rm iso}$ controls its tilt and therefore how strongly that contribution is weighted toward small scales. 
However, once blue-tilted isocurvature spectra are allowed, different combinations of amplitude and tilt can produce similar changes in the timing of the 21-cm signal. 
In particular, a larger $r_{\rm CDM}$ at fixed blue tilt and a larger $n_{\rm iso}$ at fixed amplitude can both move the characteristic features of the variance and skewness to higher redshift. 
As a result, the two parameters become difficult to disentangle and produce elongated degeneracy contours in the $(r_{\rm CDM}, n_{\rm iso})$ plane.

From an observational perspective, the forecasted  constraints suggest that, in the absence of strong systematics, SKA Phase 1 could achieve percent or sub-percent precision on both cosmological and key astrophysical parameters.

\section{Summary and Discussion}
\label{Sec4}

In this work, we have investigated the impact of CDM isocurvature perturbations on the 21cm brightness temperature signal using one-point statistics—variance and skewness—from semi-numerical simulations that include SKA-level noise and three representative astrophysical models. We find that even a small isocurvature fraction systematically advances the timing of major milestones—the onset of Wouthuysen–Field coupling, X-ray heating, and reionization—by \(\Delta z \gtrsim 1\). This shift is robust across the astrophysical models tested and is visible in both the power spectrum and one-point statistics.

Our Fisher matrix forecasts indicate that variance provides the tightest constraints in all parameters, while skewness constraints are weaker due to their higher susceptibility to thermal noise and entanglement with astrophysical parameters. Moreover, because the Fisher formalism assumes Gaussian likelihoods, it cannot fully capture the information encoded in non-Gaussian statistics; simulation-based or likelihood-free inference methods will be required to exploit the constraining power of higher-order moments \citep{2023MNRAS.524.4239P,2024ApJ...973...41Z,2025CmPhy...8..220S}.

A persistent challenge is the strong degeneracy between $r_{\rm CDM}$ and $n_{\rm iso}$: once a blue-tilted isocurvature spectrum is allowed, both parameters can enhance the small-scale power and shift the 21-cm features in similar ways. While SKA one-point statistics alone cannot fully break this degeneracy, joint analyses that incorporate higher-\(k\) probes (e.g., 21cm forest observations \citep[e.g.][]{2014PhRvD..90h3003S,2020PhRvD.101d3516S,2020PhRvD.102b3522S,2023PhRvD.107l3520S,2023NatAs...7.1116S,2025arXiv250414656S,2025MNRAS.537..364S,2025PhLB..86239342S}) and complementary astrophysical constraints (e.g., galaxy surveys or CMB isocurvature limits) can yield robust, joint constraints on both parameters.

Our analysis assumes idealized conditions, but real 21cm observations will face additional challenges from foreground contamination, calibration errors, and radio-frequency interference, all of which can bias one-point statistics if not mitigated \citep{2011PhRvD..83j3006L,2011MNRAS.413.2103P}. Addressing these systematics will require advanced mitigation strategies and robust analysis pipelines for next-generation instruments such as SKA.

Looking ahead, the 21cm signal contains a wealth of higher-order and topological information yet to be fully exploited. Statistics such as kurtosis, the bispectrum, or persistent homology can probe non-Gaussian features and the evolving morphology of reionization\citep[e.g.][]{2016MNRAS.458.3003S,2017MNRAS.468.1542S,2018MNRAS.476.4007M,2019MNRAS.482.2653W,2020MNRAS.492..653H,2017MNRAS.465..394Y,2019ApJ...885...23C,2021MNRAS.505.1863G}. Machine learning and simulation-based inference offer further opportunities for extracting hidden patterns and constraining complex astrophysical–cosmological models \citep{2025CmPhy...8..220S,2022ApJ...926..151Z,2019MNRAS.484..282G}. 

Finally, the methodology and approaches demonstrated here can be readily extended to explore a broader range of early-universe physics beyond isocurvature, including the search for primordial non-Gaussianity, signatures of primordial black holes, exotic dark matter scenarios, or other departures from standard inflationary predictions. As a result, precision 21cm measurement becomes a powerful and versatile probe of fundamental physics in the coming decade \citep{2012RPPh...75h6901P,2006PhR...433..181F,2018PhRvD..98d3006C,2015PhRvD..92h3508M}.

\begin{acknowledgments}

This work is supported by the National SKA Program of China (No.2020SKA0110401), NSFC (Grant No.~12103044), and Yunnan Provincial Key Laboratory of Survey Science with project No. 202449CE340002. We appreciate Teppei Minoda for his valuable comments.

\end{acknowledgments}


\appendix

\section{CMB-motivated benchmark for the isocurvature fiducial model}
\label{app:planck_envelope}

For the uncorrelated CDI case with a free isocurvature spectral index $n_{\rm iso}$, the Planck constraints should not be interpreted as a single one-dimensional upper limit on $r_{\rm CDM}$. Instead, the allowed range depends jointly on the amplitude and spectral tilt of the isocurvature component. We therefore construct a conservative CMB-motivated benchmark envelope in the $(r_{\rm CDM},n_{\rm iso})$ plane.

In this construction, we use the published upper limits on the isocurvature fraction at several reference scales and translate them into equivalent upper limits on the pivot-scale ratio $r_{\rm CDM}$, following the parametrization introduced in Sec.~II. For each value of $n_{\rm iso}$, we then take the most restrictive of these limits to define a conservative upper envelope, $r_{\max}(n_{\rm iso})$.

\begin{figure}[htbp]
    \centering
    \includegraphics[width=0.48\textwidth]{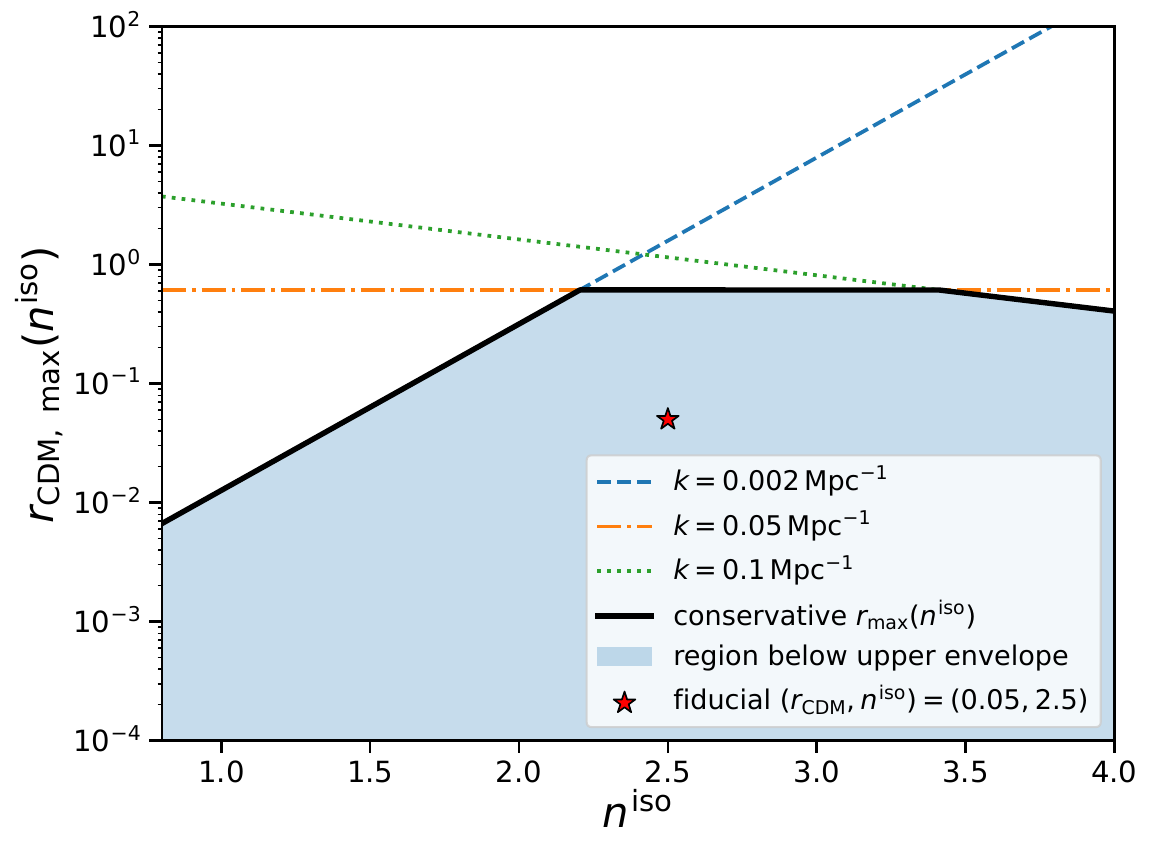}
    \caption{
    Conservative CMB-motivated upper envelope for the pivot-scale isocurvature ratio $r_{\rm CDM}$ as a function of $n_{\rm iso}$.
    The dashed, dash-dotted, and dotted curves show the upper limits inferred from the isocurvature-fraction bounds at $k=0.002$, $0.05$, and $0.1\,{\rm Mpc}^{-1}$, respectively.
    The solid black curve shows the combined conservative envelope, $r_{\max}(n_{\rm iso})$, obtained by taking the most restrictive limit at each value of $n_{\rm iso}$.
    The shaded region indicates the parameter space below this upper envelope.
    The red star marks the fiducial benchmark used in this work, $(r_{\rm CDM},n_{\rm iso})=(0.05,2.5)$, which lies inside the conservative allowed region.
    }
    \label{fig:planck_envelope}
\end{figure}

This procedure is not intended to reconstruct the full Planck posterior or the correlations among the constraints at different reference scales. Rather, it provides a conservative CMB-motivated benchmark for selecting representative 21cm simulation models. As shown in Fig.~\ref{fig:planck_envelope}, the fiducial point adopted in our Fisher analysis, $(r_{\rm CDM},n_{\rm iso})=(0.05,2.5)$, lies below this envelope. We therefore use it as a representative model that remains compatible with this conservative CMB-motivated criterion while still producing a visible response in the 21cm observables.

\section{Interplay between astrophysical models and isocurvature perturbations}
\label{app:interplay_astro_rcdm}

In this appendix, we examine the combined effect of astrophysical assumptions and isocurvature perturbations. This comparison is useful because variations in astrophysical parameters and variations in the isocurvature component can both shift the timing of major 21cm features. It is therefore important to check whether their effects can be regarded as independent, or whether they are coupled through the redshift evolution of the signal.

\begin{figure*}[htbp]
    \centering
    \includegraphics[width=0.98\textwidth]{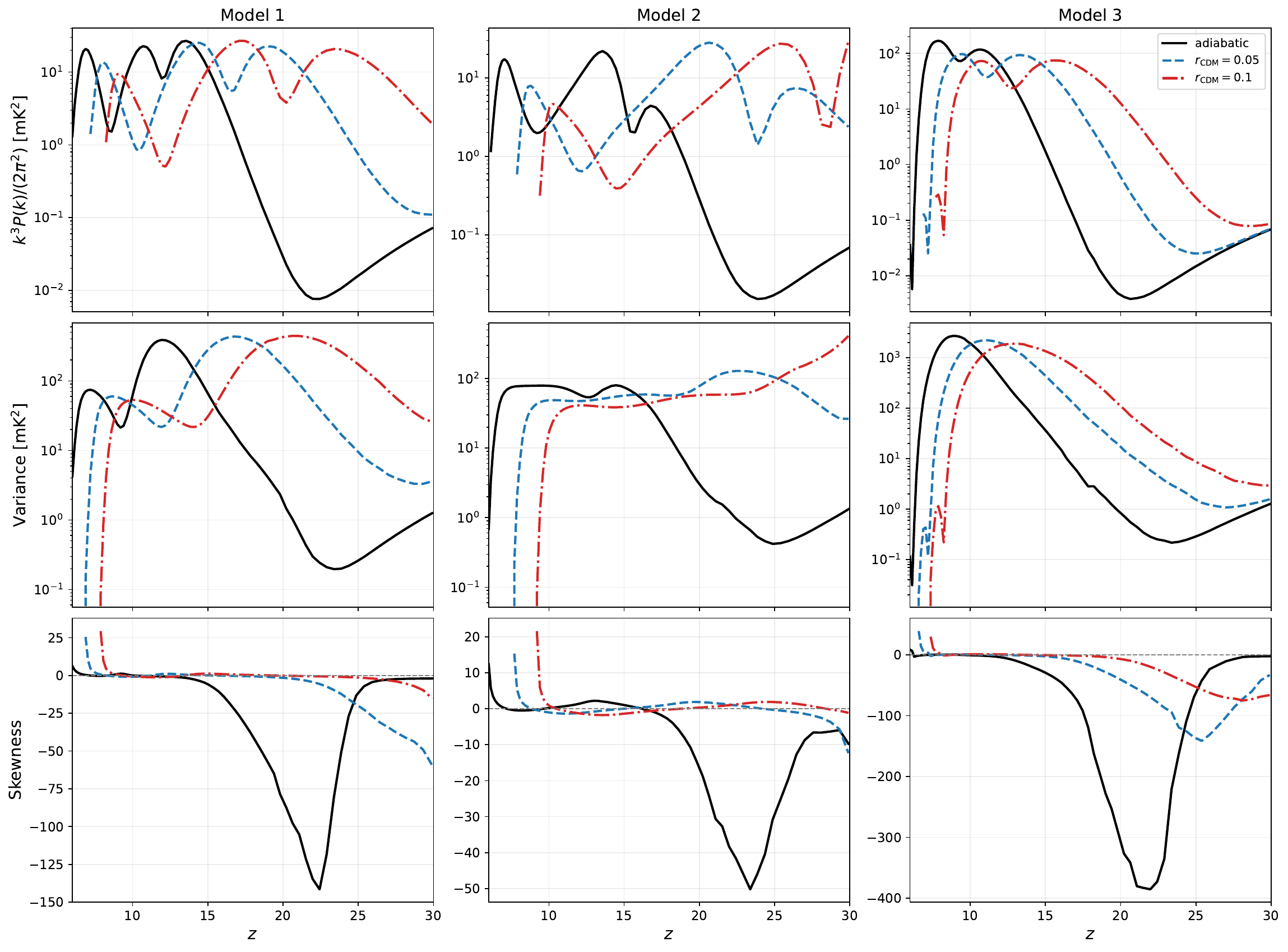}
    \caption{
    Interplay between astrophysical models and the isocurvature fraction $r_{\rm CDM}$.
    The three columns correspond to the three astrophysical models listed in Table~I.
    The top, middle, and bottom rows show the redshift evolution of the 21cm power spectrum at $k=0.1\,{\rm Mpc}^{-1}$, the variance, and the skewness, respectively.
    In each panel, we compare the adiabatic case with the isocurvature cases $r_{\rm CDM}=0.05$ and $r_{\rm CDM}=0.1$, fixing $n_{\rm iso}=2.5$.
    }
    \label{fig:app_interplay_astro_rcdm}
\end{figure*}

Figure~\ref{fig:app_interplay_astro_rcdm} compares the redshift evolution of the 21cm power spectrum, variance, and skewness for the three astrophysical models listed in Table~I. In each model, we compare the adiabatic case with two isocurvature cases, $r_{\rm CDM}=0.05$ and $r_{\rm CDM}=0.1$, while fixing $n_{\rm iso}=2.5$.

Across all three astrophysical models, increasing $r_{\rm CDM}$ generally shifts the characteristic features of the observables toward higher redshift. This trend reflects the earlier formation of small-scale structures caused by the enhanced small-scale power from the blue-tilted isocurvature component. The qualitative timing shift induced by isocurvature perturbations is therefore robust against the choice of astrophysical model.

At the same time, the detailed amplitudes, widths, and shapes of the features remain strongly model dependent. The astrophysical parameters control the efficiency and timing of star formation, X-ray heating, and reionization, and therefore determine the detailed redshift evolution of the power-spectrum peaks and one-point statistics. Thus, the combined effect is not a simple additive offset applied to a fixed astrophysical background. Instead, the isocurvature component mainly changes the timing of the main features, while the astrophysical model controls their amplitudes and morphology.

This comparison motivates the full marginalized Fisher analysis in the main text, where the isocurvature parameters are constrained simultaneously with the astrophysical parameters rather than being varied in isolation.

\section{Dependence on the smoothing scale}
\label{app:smoothing}

The smoothing scale is important because the native simulation resolution is much finer than the effective angular resolution relevant for SKA-like observations. Smoothing suppresses small-scale brightness-temperature fluctuations and reduces pixel-level thermal noise, but it can also remove part of the small-scale information that is potentially sensitive to isocurvature perturbations. We therefore examine how the one-point statistics depend on the adopted smoothing scale.

\begin{figure*}[htbp]
    \centering
    \includegraphics[width=0.68\textwidth]{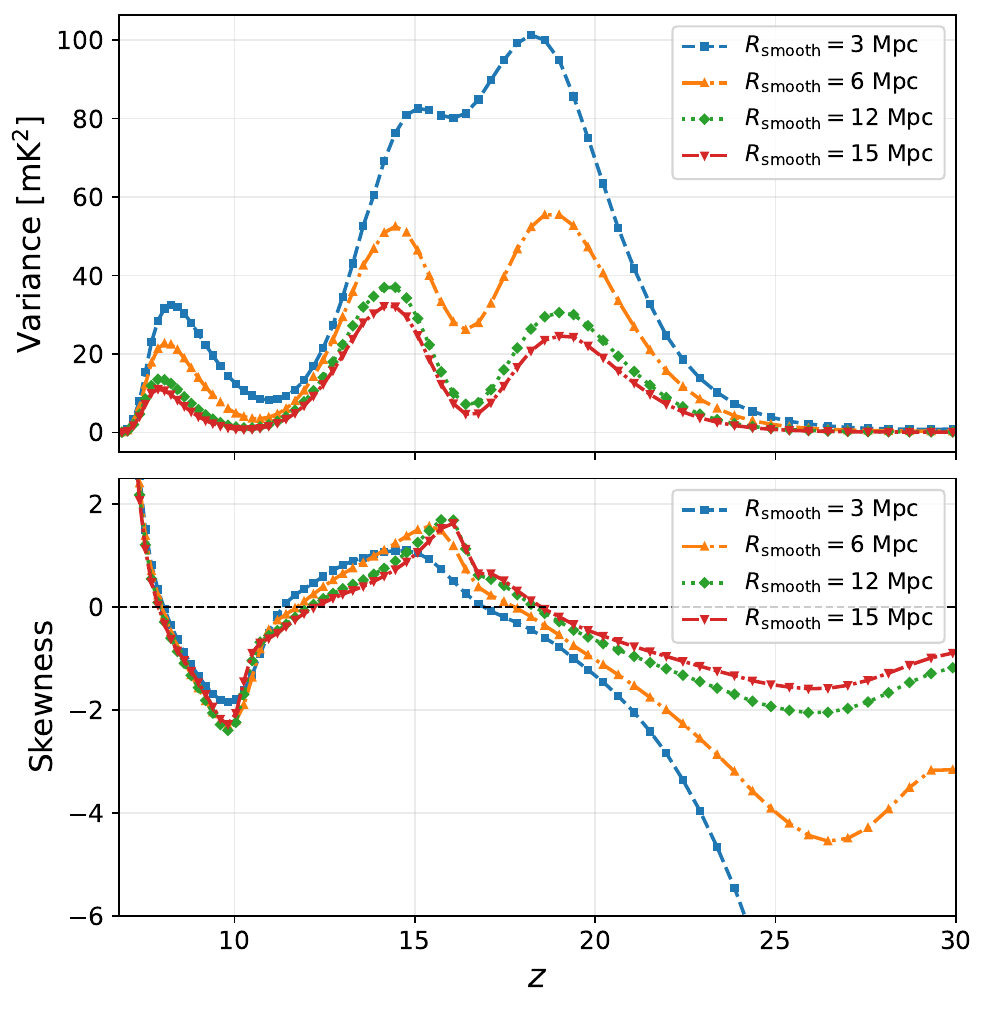}
    \caption{
    Redshift evolution of the one-point statistics of the 21cm brightness temperature for the isocurvature case with $r_{\rm CDM}=0.05$ and $n_{\rm iso}=2.5$, adopting astrophysical model~1 in Table~I.
    The top and bottom panels show the variance and skewness, respectively, for different smoothing scales, $R_{\rm smooth}=3,\ 6,\ 12,\ 15$ Mpc.
    }
    \label{fig:appendix_smoothing_scan}
\end{figure*}

Figure~\ref{fig:appendix_smoothing_scan} shows the redshift evolution of the variance and skewness for $R_{\rm smooth}=3,\ 6,\ 12,\ 15$ Mpc, adopting the fiducial isocurvature model with $r_{\rm CDM}=0.05$ and $n_{\rm iso}=2.5$. As expected, increasing the smoothing scale suppresses the amplitude of the variance because small-scale fluctuations are averaged out. The skewness is also affected, especially in redshift ranges where the brightness-temperature distribution has extended tails.

Nevertheless, the qualitative redshift evolution remains stable. In particular, the locations of the main variance features and the broad evolution of the skewness are not qualitatively altered by the smoothing scale. This indicates that the timing-based signatures discussed in the main text are not artifacts of a particular smoothing choice. The fiducial value $R_{\rm smooth}=12$ Mpc used in the main analysis should therefore be regarded as an observationally motivated smoothing scale rather than as a fine-tuned parameter.

\begingroup
\sloppy
\emergencystretch=3em
\bibliography{ref}

\begin{thebibliography}{90}
\expandafter\ifx\csname natexlab\endcsname\relax\def\natexlab#1{#1}\fi
\expandafter\ifx\csname bibnamefont\endcsname\relax
  \def\bibnamefont#1{#1}\fi
\expandafter\ifx\csname bibfnamefont\endcsname\relax
  \def\bibfnamefont#1{#1}\fi
\expandafter\ifx\csname citenamefont\endcsname\relax
  \def\citenamefont#1{#1}\fi
\expandafter\ifx\csname url\endcsname\relax
  \def\url#1{\texttt{#1}}\fi
\expandafter\ifx\csname urlprefix\endcsname\relax\def\urlprefix{URL }\fi
\providecommand{\bibinfo}[2]{#2}
\providecommand{\eprint}[2][]{\url{#2}}

\bibitem[{\citenamefont{{Planck Collaboration} et~al.}(2020{\natexlab{a}})\citenamefont{{Planck Collaboration}, {Akrami}, {Arroja}, {Ashdown}, {Aumont}, {Baccigalupi}, {Ballardini}, {Banday}, {Barreiro}, {Bartolo} et~al.}}]{2020A&A...641A..10P}
\bibinfo{author}{\bibnamefont{{Planck Collaboration}}}, \bibinfo{author}{\bibfnamefont{Y.}~\bibnamefont{{Akrami}}}, \bibinfo{author}{\bibfnamefont{F.}~\bibnamefont{{Arroja}}}, \bibinfo{author}{\bibfnamefont{M.}~\bibnamefont{{Ashdown}}}, \bibinfo{author}{\bibfnamefont{J.}~\bibnamefont{{Aumont}}}, \bibinfo{author}{\bibfnamefont{C.}~\bibnamefont{{Baccigalupi}}}, \bibinfo{author}{\bibfnamefont{M.}~\bibnamefont{{Ballardini}}}, \bibinfo{author}{\bibfnamefont{A.~J.} \bibnamefont{{Banday}}}, \bibinfo{author}{\bibfnamefont{R.~B.} \bibnamefont{{Barreiro}}}, \bibinfo{author}{\bibfnamefont{N.}~\bibnamefont{{Bartolo}}}, \bibnamefont{et~al.}, \bibinfo{journal}{\aap} \textbf{\bibinfo{volume}{641}}, \bibinfo{eid}{A10} (\bibinfo{year}{2020}{\natexlab{a}}), \eprint{1807.06211}.

\bibitem[{\citenamefont{{Komatsu} et~al.}(2011)\citenamefont{{Komatsu}, {Smith}, {Dunkley}, {Bennett}, {Gold}, {Hinshaw}, {Jarosik}, {Larson}, {Nolta}, {Page} et~al.}}]{2011ApJS..192...18K}
\bibinfo{author}{\bibfnamefont{E.}~\bibnamefont{{Komatsu}}}, \bibinfo{author}{\bibfnamefont{K.~M.} \bibnamefont{{Smith}}}, \bibinfo{author}{\bibfnamefont{J.}~\bibnamefont{{Dunkley}}}, \bibinfo{author}{\bibfnamefont{C.~L.} \bibnamefont{{Bennett}}}, \bibinfo{author}{\bibfnamefont{B.}~\bibnamefont{{Gold}}}, \bibinfo{author}{\bibfnamefont{G.}~\bibnamefont{{Hinshaw}}}, \bibinfo{author}{\bibfnamefont{N.}~\bibnamefont{{Jarosik}}}, \bibinfo{author}{\bibfnamefont{D.}~\bibnamefont{{Larson}}}, \bibinfo{author}{\bibfnamefont{M.~R.} \bibnamefont{{Nolta}}}, \bibinfo{author}{\bibfnamefont{L.}~\bibnamefont{{Page}}}, \bibnamefont{et~al.}, \bibinfo{journal}{\apjs} \textbf{\bibinfo{volume}{192}}, \bibinfo{eid}{18} (\bibinfo{year}{2011}), \eprint{1001.4538}.

\bibitem[{\citenamefont{Bucher et~al.}(2000)\citenamefont{Bucher, Moodley, and Turok}}]{PhysRevD.62.083508}
\bibinfo{author}{\bibfnamefont{M.}~\bibnamefont{Bucher}}, \bibinfo{author}{\bibfnamefont{K.}~\bibnamefont{Moodley}}, \bibnamefont{and} \bibinfo{author}{\bibfnamefont{N.}~\bibnamefont{Turok}}, \bibinfo{journal}{Phys. Rev. D} \textbf{\bibinfo{volume}{62}}, \bibinfo{pages}{083508} (\bibinfo{year}{2000}), \urlprefix\url{https://link.aps.org/doi/10.1103/PhysRevD.62.083508}.

\bibitem[{\citenamefont{{V{\"a}liviita} and {Giannantonio}}(2009)}]{2009PhRvD..80l3516V}
\bibinfo{author}{\bibfnamefont{J.}~\bibnamefont{{V{\"a}liviita}}} \bibnamefont{and} \bibinfo{author}{\bibfnamefont{T.}~\bibnamefont{{Giannantonio}}}, \bibinfo{journal}{\prd} \textbf{\bibinfo{volume}{80}}, \bibinfo{eid}{123516} (\bibinfo{year}{2009}), \eprint{0909.5190}.

\bibitem[{\citenamefont{Kasuya and Kawasaki}(2009)}]{Kasuya_2009}
\bibinfo{author}{\bibfnamefont{S.}~\bibnamefont{Kasuya}} \bibnamefont{and} \bibinfo{author}{\bibfnamefont{M.}~\bibnamefont{Kawasaki}}, \bibinfo{journal}{Physical Review D} \textbf{\bibinfo{volume}{80}} (\bibinfo{year}{2009}), ISSN \bibinfo{issn}{1550-2368}, \urlprefix\url{http://dx.doi.org/10.1103/PhysRevD.80.023516}.

\bibitem[{\citenamefont{Chung and Yoo}(2015)}]{Chung_2015}
\bibinfo{author}{\bibfnamefont{D.~J.} \bibnamefont{Chung}} \bibnamefont{and} \bibinfo{author}{\bibfnamefont{H.}~\bibnamefont{Yoo}}, \bibinfo{journal}{Physical Review D} \textbf{\bibinfo{volume}{91}} (\bibinfo{year}{2015}), ISSN \bibinfo{issn}{1550-2368}, \urlprefix\url{http://dx.doi.org/10.1103/PhysRevD.91.083530}.

\bibitem[{\citenamefont{Chung and Upadhye}(2018)}]{Chung_2018}
\bibinfo{author}{\bibfnamefont{D.~J.} \bibnamefont{Chung}} \bibnamefont{and} \bibinfo{author}{\bibfnamefont{A.}~\bibnamefont{Upadhye}}, \bibinfo{journal}{Physical Review D} \textbf{\bibinfo{volume}{98}} (\bibinfo{year}{2018}), ISSN \bibinfo{issn}{2470-0029}, \urlprefix\url{http://dx.doi.org/10.1103/PhysRevD.98.023525}.

\bibitem[{\citenamefont{Chung and Tadepalli}(2022)}]{Chung_2022}
\bibinfo{author}{\bibfnamefont{D.~J.} \bibnamefont{Chung}} \bibnamefont{and} \bibinfo{author}{\bibfnamefont{S.~C.} \bibnamefont{Tadepalli}}, \bibinfo{journal}{Physical Review D} \textbf{\bibinfo{volume}{105}} (\bibinfo{year}{2022}), ISSN \bibinfo{issn}{2470-0029}, \urlprefix\url{http://dx.doi.org/10.1103/PhysRevD.105.123511}.

\bibitem[{\citenamefont{Afshordi et~al.}(2003)\citenamefont{Afshordi, McDonald, and Spergel}}]{Afshordi_2003}
\bibinfo{author}{\bibfnamefont{N.}~\bibnamefont{Afshordi}}, \bibinfo{author}{\bibfnamefont{P.}~\bibnamefont{McDonald}}, \bibnamefont{and} \bibinfo{author}{\bibfnamefont{D.~N.} \bibnamefont{Spergel}}, \bibinfo{journal}{The Astrophysical Journal} \textbf{\bibinfo{volume}{594}}, \bibinfo{pages}{L71–L74} (\bibinfo{year}{2003}), ISSN \bibinfo{issn}{1538-4357}, \urlprefix\url{http://dx.doi.org/10.1086/378763}.

\bibitem[{\citenamefont{Kashlinsky}(2016)}]{Kashlinsky_2016}
\bibinfo{author}{\bibfnamefont{A.}~\bibnamefont{Kashlinsky}}, \bibinfo{journal}{The Astrophysical Journal Letters} \textbf{\bibinfo{volume}{823}}, \bibinfo{pages}{L25} (\bibinfo{year}{2016}), ISSN \bibinfo{issn}{2041-8213}, \urlprefix\url{http://dx.doi.org/10.3847/2041-8205/823/2/L25}.

\bibitem[{\citenamefont{Gong and Kitajima}(2017)}]{Gong_2017}
\bibinfo{author}{\bibfnamefont{J.-O.} \bibnamefont{Gong}} \bibnamefont{and} \bibinfo{author}{\bibfnamefont{N.}~\bibnamefont{Kitajima}}, \bibinfo{journal}{Journal of Cosmology and Astroparticle Physics} \textbf{\bibinfo{volume}{2017}}, \bibinfo{pages}{017–017} (\bibinfo{year}{2017}), ISSN \bibinfo{issn}{1475-7516}, \urlprefix\url{http://dx.doi.org/10.1088/1475-7516/2017/08/017}.

\bibitem[{\citenamefont{Gong and Kitajima}(2018)}]{Gong_2018}
\bibinfo{author}{\bibfnamefont{J.-O.} \bibnamefont{Gong}} \bibnamefont{and} \bibinfo{author}{\bibfnamefont{N.}~\bibnamefont{Kitajima}}, \bibinfo{journal}{Journal of Cosmology and Astroparticle Physics} \textbf{\bibinfo{volume}{2018}}, \bibinfo{pages}{041–041} (\bibinfo{year}{2018}), ISSN \bibinfo{issn}{1475-7516}, \urlprefix\url{http://dx.doi.org/10.1088/1475-7516/2018/11/041}.

\bibitem[{\citenamefont{Mena et~al.}(2019)\citenamefont{Mena, Palomares-Ruiz, Villanueva-Domingo, and Witte}}]{Mena_2019}
\bibinfo{author}{\bibfnamefont{O.}~\bibnamefont{Mena}}, \bibinfo{author}{\bibfnamefont{S.}~\bibnamefont{Palomares-Ruiz}}, \bibinfo{author}{\bibfnamefont{P.}~\bibnamefont{Villanueva-Domingo}}, \bibnamefont{and} \bibinfo{author}{\bibfnamefont{S.~J.} \bibnamefont{Witte}}, \bibinfo{journal}{Physical Review D} \textbf{\bibinfo{volume}{100}} (\bibinfo{year}{2019}), ISSN \bibinfo{issn}{2470-0029}, \urlprefix\url{http://dx.doi.org/10.1103/PhysRevD.100.043540}.

\bibitem[{\citenamefont{Tashiro and Kadota}(2021)}]{Tashiro_2021}
\bibinfo{author}{\bibfnamefont{H.}~\bibnamefont{Tashiro}} \bibnamefont{and} \bibinfo{author}{\bibfnamefont{K.}~\bibnamefont{Kadota}}, \bibinfo{journal}{Physical Review D} \textbf{\bibinfo{volume}{104}} (\bibinfo{year}{2021}), ISSN \bibinfo{issn}{2470-0029}, \urlprefix\url{http://dx.doi.org/10.1103/PhysRevD.104.063522}.

\bibitem[{\citenamefont{{Liddle} and {Mazumdar}}(2000)}]{2000PhRvD..61l3507L}
\bibinfo{author}{\bibfnamefont{A.~R.} \bibnamefont{{Liddle}}} \bibnamefont{and} \bibinfo{author}{\bibfnamefont{A.}~\bibnamefont{{Mazumdar}}}, \bibinfo{journal}{\prd} \textbf{\bibinfo{volume}{61}}, \bibinfo{eid}{123507} (\bibinfo{year}{2000}), \eprint{astro-ph/9912349}.

\bibitem[{\citenamefont{Gordon et~al.}(2000)\citenamefont{Gordon, Wands, Bassett, and Maartens}}]{2000Adiabatic}
\bibinfo{author}{\bibfnamefont{C.}~\bibnamefont{Gordon}}, \bibinfo{author}{\bibfnamefont{D.}~\bibnamefont{Wands}}, \bibinfo{author}{\bibfnamefont{B.~A.} \bibnamefont{Bassett}}, \bibnamefont{and} \bibinfo{author}{\bibfnamefont{R.}~\bibnamefont{Maartens}}, \bibinfo{journal}{Physical review D: Particles and fields} \textbf{\bibinfo{volume}{63}}, \bibinfo{pages}{398} (\bibinfo{year}{2000}).

\bibitem[{\citenamefont{{Pritchard} and {Loeb}}(2012)}]{2012RPPh...75h6901P}
\bibinfo{author}{\bibfnamefont{J.~R.} \bibnamefont{{Pritchard}}} \bibnamefont{and} \bibinfo{author}{\bibfnamefont{A.}~\bibnamefont{{Loeb}}}, \bibinfo{journal}{Reports on Progress in Physics} \textbf{\bibinfo{volume}{75}}, \bibinfo{eid}{086901} (\bibinfo{year}{2012}), \eprint{1109.6012}.

\bibitem[{\citenamefont{{Hasegawa} et~al.}(2016)\citenamefont{{Hasegawa}, {Asaba}, {Ichiki}, {Inoue}, {Inoue}, {Ishiyama}, {Shimabukuro}, {Takahashi}, {Tashiro}, {Yajima} et~al.}}]{2016arXiv160301961H}
\bibinfo{author}{\bibfnamefont{K.}~\bibnamefont{{Hasegawa}}}, \bibinfo{author}{\bibfnamefont{S.}~\bibnamefont{{Asaba}}}, \bibinfo{author}{\bibfnamefont{K.}~\bibnamefont{{Ichiki}}}, \bibinfo{author}{\bibfnamefont{A.~K.} \bibnamefont{{Inoue}}}, \bibinfo{author}{\bibfnamefont{S.}~\bibnamefont{{Inoue}}}, \bibinfo{author}{\bibfnamefont{T.}~\bibnamefont{{Ishiyama}}}, \bibinfo{author}{\bibfnamefont{H.}~\bibnamefont{{Shimabukuro}}}, \bibinfo{author}{\bibfnamefont{K.}~\bibnamefont{{Takahashi}}}, \bibinfo{author}{\bibfnamefont{H.}~\bibnamefont{{Tashiro}}}, \bibinfo{author}{\bibfnamefont{H.}~\bibnamefont{{Yajima}}}, \bibnamefont{et~al.}, \bibinfo{journal}{arXiv e-prints} \bibinfo{eid}{arXiv:1603.01961} (\bibinfo{year}{2016}), \eprint{1603.01961}.

\bibitem[{\citenamefont{{Yamauchi} et~al.}(2016)\citenamefont{{Yamauchi}, {Ichiki}, {Kohri}, {Namikawa}, {Oyama}, {Sekiguchi}, {Shimabukuro}, {Takahashi}, {Takahashi}, {Yokoyama} et~al.}}]{2016PASJ...68R...2Y}
\bibinfo{author}{\bibfnamefont{D.}~\bibnamefont{{Yamauchi}}}, \bibinfo{author}{\bibfnamefont{K.}~\bibnamefont{{Ichiki}}}, \bibinfo{author}{\bibfnamefont{K.}~\bibnamefont{{Kohri}}}, \bibinfo{author}{\bibfnamefont{T.}~\bibnamefont{{Namikawa}}}, \bibinfo{author}{\bibfnamefont{Y.}~\bibnamefont{{Oyama}}}, \bibinfo{author}{\bibfnamefont{T.}~\bibnamefont{{Sekiguchi}}}, \bibinfo{author}{\bibfnamefont{H.}~\bibnamefont{{Shimabukuro}}}, \bibinfo{author}{\bibfnamefont{K.}~\bibnamefont{{Takahashi}}}, \bibinfo{author}{\bibfnamefont{T.}~\bibnamefont{{Takahashi}}}, \bibinfo{author}{\bibfnamefont{S.}~\bibnamefont{{Yokoyama}}}, \bibnamefont{et~al.}, \bibinfo{journal}{\pasj} \textbf{\bibinfo{volume}{68}}, \bibinfo{eid}{R2} (\bibinfo{year}{2016}), \eprint{1603.01959}.

\bibitem[{\citenamefont{{Shimabukuro} et~al.}(2023{\natexlab{a}})\citenamefont{{Shimabukuro}, {Hasegawa}, {Kuchinomachi}, {Yajima}, and {Yoshiura}}}]{2023PASJ...75S...1S}
\bibinfo{author}{\bibfnamefont{H.}~\bibnamefont{{Shimabukuro}}}, \bibinfo{author}{\bibfnamefont{K.}~\bibnamefont{{Hasegawa}}}, \bibinfo{author}{\bibfnamefont{A.}~\bibnamefont{{Kuchinomachi}}}, \bibinfo{author}{\bibfnamefont{H.}~\bibnamefont{{Yajima}}}, \bibnamefont{and} \bibinfo{author}{\bibfnamefont{S.}~\bibnamefont{{Yoshiura}}}, \bibinfo{journal}{\pasj} \textbf{\bibinfo{volume}{75}}, \bibinfo{pages}{S1} (\bibinfo{year}{2023}{\natexlab{a}}), \eprint{2303.07594}.

\bibitem[{\citenamefont{{Minoda} et~al.}(2022)\citenamefont{{Minoda}, {Yoshiura}, and {Takahashi}}}]{2022PhRvD.105h3523M}
\bibinfo{author}{\bibfnamefont{T.}~\bibnamefont{{Minoda}}}, \bibinfo{author}{\bibfnamefont{S.}~\bibnamefont{{Yoshiura}}}, \bibnamefont{and} \bibinfo{author}{\bibfnamefont{T.}~\bibnamefont{{Takahashi}}}, \bibinfo{journal}{\prd} \textbf{\bibinfo{volume}{105}}, \bibinfo{eid}{083523} (\bibinfo{year}{2022}), \eprint{2112.15135}.

\bibitem[{\citenamefont{{Shimabukuro} et~al.}(2016)\citenamefont{{Shimabukuro}, {Yoshiura}, {Takahashi}, {Yokoyama}, and {Ichiki}}}]{2016MNRAS.458.3003S}
\bibinfo{author}{\bibfnamefont{H.}~\bibnamefont{{Shimabukuro}}}, \bibinfo{author}{\bibfnamefont{S.}~\bibnamefont{{Yoshiura}}}, \bibinfo{author}{\bibfnamefont{K.}~\bibnamefont{{Takahashi}}}, \bibinfo{author}{\bibfnamefont{S.}~\bibnamefont{{Yokoyama}}}, \bibnamefont{and} \bibinfo{author}{\bibfnamefont{K.}~\bibnamefont{{Ichiki}}}, \bibinfo{journal}{\mnras} \textbf{\bibinfo{volume}{458}}, \bibinfo{pages}{3003} (\bibinfo{year}{2016}), \eprint{1507.01335}.

\bibitem[{\citenamefont{{Kubota} et~al.}(2016)\citenamefont{{Kubota}, {Yoshiura}, {Shimabukuro}, and {Takahashi}}}]{2016PASJ...68...61K}
\bibinfo{author}{\bibfnamefont{K.}~\bibnamefont{{Kubota}}}, \bibinfo{author}{\bibfnamefont{S.}~\bibnamefont{{Yoshiura}}}, \bibinfo{author}{\bibfnamefont{H.}~\bibnamefont{{Shimabukuro}}}, \bibnamefont{and} \bibinfo{author}{\bibfnamefont{K.}~\bibnamefont{{Takahashi}}}, \bibinfo{journal}{\pasj} \textbf{\bibinfo{volume}{68}}, \bibinfo{eid}{61} (\bibinfo{year}{2016}), \eprint{1602.02873}.

\bibitem[{\citenamefont{{Shimabukuro} et~al.}(2017)\citenamefont{{Shimabukuro}, {Yoshiura}, {Takahashi}, {Yokoyama}, and {Ichiki}}}]{2017MNRAS.468.1542S}
\bibinfo{author}{\bibfnamefont{H.}~\bibnamefont{{Shimabukuro}}}, \bibinfo{author}{\bibfnamefont{S.}~\bibnamefont{{Yoshiura}}}, \bibinfo{author}{\bibfnamefont{K.}~\bibnamefont{{Takahashi}}}, \bibinfo{author}{\bibfnamefont{S.}~\bibnamefont{{Yokoyama}}}, \bibnamefont{and} \bibinfo{author}{\bibfnamefont{K.}~\bibnamefont{{Ichiki}}}, \bibinfo{journal}{\mnras} \textbf{\bibinfo{volume}{468}}, \bibinfo{pages}{1542} (\bibinfo{year}{2017}), \eprint{1608.00372}.

\bibitem[{\citenamefont{{Watkinson} and {Pritchard}}(2014)}]{2014MNRAS.443.3090W}
\bibinfo{author}{\bibfnamefont{C.~A.} \bibnamefont{{Watkinson}}} \bibnamefont{and} \bibinfo{author}{\bibfnamefont{J.~R.} \bibnamefont{{Pritchard}}}, \bibinfo{journal}{\mnras} \textbf{\bibinfo{volume}{443}}, \bibinfo{pages}{3090} (\bibinfo{year}{2014}), \eprint{1312.1342}.

\bibitem[{\citenamefont{{Shimabukuro} et~al.}(2015)\citenamefont{{Shimabukuro}, {Yoshiura}, {Takahashi}, {Yokoyama}, and {Ichiki}}}]{2015MNRAS.451..467S}
\bibinfo{author}{\bibfnamefont{H.}~\bibnamefont{{Shimabukuro}}}, \bibinfo{author}{\bibfnamefont{S.}~\bibnamefont{{Yoshiura}}}, \bibinfo{author}{\bibfnamefont{K.}~\bibnamefont{{Takahashi}}}, \bibinfo{author}{\bibfnamefont{S.}~\bibnamefont{{Yokoyama}}}, \bibnamefont{and} \bibinfo{author}{\bibfnamefont{K.}~\bibnamefont{{Ichiki}}}, \bibinfo{journal}{\mnras} \textbf{\bibinfo{volume}{451}}, \bibinfo{pages}{467} (\bibinfo{year}{2015}), \eprint{1412.3332}.

\bibitem[{\citenamefont{{Watkinson} and {Pritchard}}(2015)}]{2015MNRAS.454.1416W}
\bibinfo{author}{\bibfnamefont{C.~A.} \bibnamefont{{Watkinson}}} \bibnamefont{and} \bibinfo{author}{\bibfnamefont{J.~R.} \bibnamefont{{Pritchard}}}, \bibinfo{journal}{\mnras} \textbf{\bibinfo{volume}{454}}, \bibinfo{pages}{1416} (\bibinfo{year}{2015}), \eprint{1505.07108}.

\bibitem[{\citenamefont{{Abdurashidova} et~al.}(2022)\citenamefont{{Abdurashidova}, {Aguirre}, {Alexander}, {Ali}, {Balfour}, {Barkana}, {Beardsley}, {Bernardi}, {Billings}, {Bowman} et~al.}}]{2022ApJ...924...51A}
\bibinfo{author}{\bibfnamefont{Z.}~\bibnamefont{{Abdurashidova}}}, \bibinfo{author}{\bibfnamefont{J.~E.} \bibnamefont{{Aguirre}}}, \bibinfo{author}{\bibfnamefont{P.}~\bibnamefont{{Alexander}}}, \bibinfo{author}{\bibfnamefont{Z.~S.} \bibnamefont{{Ali}}}, \bibinfo{author}{\bibfnamefont{Y.}~\bibnamefont{{Balfour}}}, \bibinfo{author}{\bibfnamefont{R.}~\bibnamefont{{Barkana}}}, \bibinfo{author}{\bibfnamefont{A.~P.} \bibnamefont{{Beardsley}}}, \bibinfo{author}{\bibfnamefont{G.}~\bibnamefont{{Bernardi}}}, \bibinfo{author}{\bibfnamefont{T.~S.} \bibnamefont{{Billings}}}, \bibinfo{author}{\bibfnamefont{J.~D.} \bibnamefont{{Bowman}}}, \bibnamefont{et~al.}, \bibinfo{journal}{\apj} \textbf{\bibinfo{volume}{924}}, \bibinfo{eid}{51} (\bibinfo{year}{2022}), \eprint{2108.07282}.

\bibitem[{\citenamefont{{Koopmans} et~al.}(2015)\citenamefont{{Koopmans}, {Pritchard}, {Mellema}, {Aguirre}, {Ahn}, {Barkana}, {van Bemmel}, {Bernardi}, {Bonaldi}, {Briggs} et~al.}}]{2015aska.confE...1K}
\bibinfo{author}{\bibfnamefont{L.}~\bibnamefont{{Koopmans}}}, \bibinfo{author}{\bibfnamefont{J.}~\bibnamefont{{Pritchard}}}, \bibinfo{author}{\bibfnamefont{G.}~\bibnamefont{{Mellema}}}, \bibinfo{author}{\bibfnamefont{J.}~\bibnamefont{{Aguirre}}}, \bibinfo{author}{\bibfnamefont{K.}~\bibnamefont{{Ahn}}}, \bibinfo{author}{\bibfnamefont{R.}~\bibnamefont{{Barkana}}}, \bibinfo{author}{\bibfnamefont{I.}~\bibnamefont{{van Bemmel}}}, \bibinfo{author}{\bibfnamefont{G.}~\bibnamefont{{Bernardi}}}, \bibinfo{author}{\bibfnamefont{A.}~\bibnamefont{{Bonaldi}}}, \bibinfo{author}{\bibfnamefont{F.}~\bibnamefont{{Briggs}}}, \bibnamefont{et~al.} (\bibinfo{year}{2015}), p.~\bibinfo{pages}{1}, \eprint{1505.07568}.

\bibitem[{\citenamefont{{van Haarlem} et~al.}(2013)\citenamefont{{van Haarlem}, {Wise}, {Gunst}, {Heald}, {McKean}, {Hessels}, {de Bruyn}, {Nijboer}, {Swinbank}, {Fallows} et~al.}}]{2013A&A...556A...2V}
\bibinfo{author}{\bibfnamefont{M.~P.} \bibnamefont{{van Haarlem}}}, \bibinfo{author}{\bibfnamefont{M.~W.} \bibnamefont{{Wise}}}, \bibinfo{author}{\bibfnamefont{A.~W.} \bibnamefont{{Gunst}}}, \bibinfo{author}{\bibfnamefont{G.}~\bibnamefont{{Heald}}}, \bibinfo{author}{\bibfnamefont{J.~P.} \bibnamefont{{McKean}}}, \bibinfo{author}{\bibfnamefont{J.~W.~T.} \bibnamefont{{Hessels}}}, \bibinfo{author}{\bibfnamefont{A.~G.} \bibnamefont{{de Bruyn}}}, \bibinfo{author}{\bibfnamefont{R.}~\bibnamefont{{Nijboer}}}, \bibinfo{author}{\bibfnamefont{J.}~\bibnamefont{{Swinbank}}}, \bibinfo{author}{\bibfnamefont{R.}~\bibnamefont{{Fallows}}}, \bibnamefont{et~al.}, \bibinfo{journal}{\aap} \textbf{\bibinfo{volume}{556}}, \bibinfo{eid}{A2} (\bibinfo{year}{2013}), \eprint{1305.3550}.

\bibitem[{\citenamefont{{Bardeen} et~al.}(1986)\citenamefont{{Bardeen}, {Bond}, {Kaiser}, and {Szalay}}}]{1986ApJ...304...15B}
\bibinfo{author}{\bibfnamefont{J.~M.} \bibnamefont{{Bardeen}}}, \bibinfo{author}{\bibfnamefont{J.~R.} \bibnamefont{{Bond}}}, \bibinfo{author}{\bibfnamefont{N.}~\bibnamefont{{Kaiser}}}, \bibnamefont{and} \bibinfo{author}{\bibfnamefont{A.~S.} \bibnamefont{{Szalay}}}, \bibinfo{journal}{\apj} \textbf{\bibinfo{volume}{304}}, \bibinfo{pages}{15} (\bibinfo{year}{1986}).

\bibitem[{\citenamefont{{Sugiyama}}(1995)}]{1995ApJS..100..281S}
\bibinfo{author}{\bibfnamefont{N.}~\bibnamefont{{Sugiyama}}}, \bibinfo{journal}{\apjs} \textbf{\bibinfo{volume}{100}}, \bibinfo{pages}{281} (\bibinfo{year}{1995}), \eprint{astro-ph/9412025}.

\bibitem[{\citenamefont{{Planck Collaboration} et~al.}(2016{\natexlab{a}})\citenamefont{{Planck Collaboration}, {Ade}, {Aghanim}, {Arnaud}, {Ashdown}, {Aumont}, {Baccigalupi}, {Banday}, {Barreiro}, {Bartlett} et~al.}}]{2016A&A...594A..13P}
\bibinfo{author}{\bibnamefont{{Planck Collaboration}}}, \bibinfo{author}{\bibfnamefont{P.~A.~R.} \bibnamefont{{Ade}}}, \bibinfo{author}{\bibfnamefont{N.}~\bibnamefont{{Aghanim}}}, \bibinfo{author}{\bibfnamefont{M.}~\bibnamefont{{Arnaud}}}, \bibinfo{author}{\bibfnamefont{M.}~\bibnamefont{{Ashdown}}}, \bibinfo{author}{\bibfnamefont{J.}~\bibnamefont{{Aumont}}}, \bibinfo{author}{\bibfnamefont{C.}~\bibnamefont{{Baccigalupi}}}, \bibinfo{author}{\bibfnamefont{A.~J.} \bibnamefont{{Banday}}}, \bibinfo{author}{\bibfnamefont{R.~B.} \bibnamefont{{Barreiro}}}, \bibinfo{author}{\bibfnamefont{J.~G.} \bibnamefont{{Bartlett}}}, \bibnamefont{et~al.}, \bibinfo{journal}{\aap} \textbf{\bibinfo{volume}{594}}, \bibinfo{eid}{A13} (\bibinfo{year}{2016}{\natexlab{a}}), \eprint{1502.01589}.

\bibitem[{\citenamefont{{Planck Collaboration} et~al.}(2016{\natexlab{b}})\citenamefont{{Planck Collaboration}, {Ade}, {Aghanim}, {Arnaud}, {Arroja}, {Ashdown}, {Aumont}, {Baccigalupi}, {Ballardini}, {Banday} et~al.}}]{2016A&A...594A..17P}
\bibinfo{author}{\bibnamefont{{Planck Collaboration}}}, \bibinfo{author}{\bibfnamefont{P.~A.~R.} \bibnamefont{{Ade}}}, \bibinfo{author}{\bibfnamefont{N.}~\bibnamefont{{Aghanim}}}, \bibinfo{author}{\bibfnamefont{M.}~\bibnamefont{{Arnaud}}}, \bibinfo{author}{\bibfnamefont{F.}~\bibnamefont{{Arroja}}}, \bibinfo{author}{\bibfnamefont{M.}~\bibnamefont{{Ashdown}}}, \bibinfo{author}{\bibfnamefont{J.}~\bibnamefont{{Aumont}}}, \bibinfo{author}{\bibfnamefont{C.}~\bibnamefont{{Baccigalupi}}}, \bibinfo{author}{\bibfnamefont{M.}~\bibnamefont{{Ballardini}}}, \bibinfo{author}{\bibfnamefont{A.~J.} \bibnamefont{{Banday}}}, \bibnamefont{et~al.}, \bibinfo{journal}{\aap} \textbf{\bibinfo{volume}{594}}, \bibinfo{eid}{A17} (\bibinfo{year}{2016}{\natexlab{b}}), \eprint{1502.01592}.

\bibitem[{buc(2025)}]{buckley2025generalconstraintsisocurvaturecmb}
 (\bibinfo{year}{2025}), \eprint{2502.20434}, \urlprefix\url{https://arxiv.org/abs/2502.20434}.

\bibitem[{\citenamefont{{Umeda} et~al.}(2024)\citenamefont{{Umeda}, {Ouchi}, {Kikuta}, {Harikane}, {Ono}, {Shibuya}, {Inoue}, {Shimasaku}, {Liang}, {Matsumoto} et~al.}}]{2024arXiv241115495U}
\bibinfo{author}{\bibfnamefont{H.}~\bibnamefont{{Umeda}}}, \bibinfo{author}{\bibfnamefont{M.}~\bibnamefont{{Ouchi}}}, \bibinfo{author}{\bibfnamefont{S.}~\bibnamefont{{Kikuta}}}, \bibinfo{author}{\bibfnamefont{Y.}~\bibnamefont{{Harikane}}}, \bibinfo{author}{\bibfnamefont{Y.}~\bibnamefont{{Ono}}}, \bibinfo{author}{\bibfnamefont{T.}~\bibnamefont{{Shibuya}}}, \bibinfo{author}{\bibfnamefont{A.~K.} \bibnamefont{{Inoue}}}, \bibinfo{author}{\bibfnamefont{K.}~\bibnamefont{{Shimasaku}}}, \bibinfo{author}{\bibfnamefont{Y.}~\bibnamefont{{Liang}}}, \bibinfo{author}{\bibfnamefont{A.}~\bibnamefont{{Matsumoto}}}, \bibnamefont{et~al.}, \bibinfo{journal}{arXiv e-prints} \bibinfo{eid}{arXiv:2411.15495} (\bibinfo{year}{2024}), \eprint{2411.15495}.

\bibitem[{\citenamefont{Ouchi et~al.}(2010)\citenamefont{Ouchi, Shimasaku, Furusawa, Saito, Yoshida, Akiyama, Ono, Yamada, Ota, Kashikawa et~al.}}]{Ouchi_2010}
\bibinfo{author}{\bibfnamefont{M.}~\bibnamefont{Ouchi}}, \bibinfo{author}{\bibfnamefont{K.}~\bibnamefont{Shimasaku}}, \bibinfo{author}{\bibfnamefont{H.}~\bibnamefont{Furusawa}}, \bibinfo{author}{\bibfnamefont{T.}~\bibnamefont{Saito}}, \bibinfo{author}{\bibfnamefont{M.}~\bibnamefont{Yoshida}}, \bibinfo{author}{\bibfnamefont{M.}~\bibnamefont{Akiyama}}, \bibinfo{author}{\bibfnamefont{Y.}~\bibnamefont{Ono}}, \bibinfo{author}{\bibfnamefont{T.}~\bibnamefont{Yamada}}, \bibinfo{author}{\bibfnamefont{K.}~\bibnamefont{Ota}}, \bibinfo{author}{\bibfnamefont{N.}~\bibnamefont{Kashikawa}}, \bibnamefont{et~al.}, \bibinfo{journal}{The Astrophysical Journal} \textbf{\bibinfo{volume}{723}}, \bibinfo{pages}{869} (\bibinfo{year}{2010}), \urlprefix\url{https://dx.doi.org/10.1088/0004-637X/723/1/869}.

\bibitem[{\citenamefont{Konno et~al.}(2014)\citenamefont{Konno, Ouchi, Ono, Shimasaku, Shibuya, Furusawa, Nakajima, Naito, Momose, Yuma et~al.}}]{Konno_2014}
\bibinfo{author}{\bibfnamefont{A.}~\bibnamefont{Konno}}, \bibinfo{author}{\bibfnamefont{M.}~\bibnamefont{Ouchi}}, \bibinfo{author}{\bibfnamefont{Y.}~\bibnamefont{Ono}}, \bibinfo{author}{\bibfnamefont{K.}~\bibnamefont{Shimasaku}}, \bibinfo{author}{\bibfnamefont{T.}~\bibnamefont{Shibuya}}, \bibinfo{author}{\bibfnamefont{H.}~\bibnamefont{Furusawa}}, \bibinfo{author}{\bibfnamefont{K.}~\bibnamefont{Nakajima}}, \bibinfo{author}{\bibfnamefont{Y.}~\bibnamefont{Naito}}, \bibinfo{author}{\bibfnamefont{R.}~\bibnamefont{Momose}}, \bibinfo{author}{\bibfnamefont{S.}~\bibnamefont{Yuma}}, \bibnamefont{et~al.}, \bibinfo{journal}{The Astrophysical Journal} \textbf{\bibinfo{volume}{797}}, \bibinfo{pages}{16} (\bibinfo{year}{2014}), \urlprefix\url{https://dx.doi.org/10.1088/0004-637X/797/1/16}.

\bibitem[{\citenamefont{Zheng et~al.}(2017)\citenamefont{Zheng, Wang, Rhoads, Infante, Malhotra, Hu, Walker, Jiang, Jiang, Hibon et~al.}}]{Zheng_2017}
\bibinfo{author}{\bibfnamefont{Z.-Y.} \bibnamefont{Zheng}}, \bibinfo{author}{\bibfnamefont{J.}~\bibnamefont{Wang}}, \bibinfo{author}{\bibfnamefont{J.}~\bibnamefont{Rhoads}}, \bibinfo{author}{\bibfnamefont{L.}~\bibnamefont{Infante}}, \bibinfo{author}{\bibfnamefont{S.}~\bibnamefont{Malhotra}}, \bibinfo{author}{\bibfnamefont{W.}~\bibnamefont{Hu}}, \bibinfo{author}{\bibfnamefont{A.~R.} \bibnamefont{Walker}}, \bibinfo{author}{\bibfnamefont{L.}~\bibnamefont{Jiang}}, \bibinfo{author}{\bibfnamefont{C.}~\bibnamefont{Jiang}}, \bibinfo{author}{\bibfnamefont{P.}~\bibnamefont{Hibon}}, \bibnamefont{et~al.}, \bibinfo{journal}{The Astrophysical Journal Letters} \textbf{\bibinfo{volume}{842}}, \bibinfo{pages}{L22} (\bibinfo{year}{2017}), \urlprefix\url{https://dx.doi.org/10.3847/2041-8213/aa794f}.

\bibitem[{\citenamefont{{Inoue} et~al.}(2018)\citenamefont{{Inoue}, {Hasegawa}, {Ishiyama}, {Yajima}, {Shimizu}, {Umemura}, {Konno}, {Harikane}, {Shibuya}, {Ouchi} et~al.}}]{2018PASJ...70...55I}
\bibinfo{author}{\bibfnamefont{A.~K.} \bibnamefont{{Inoue}}}, \bibinfo{author}{\bibfnamefont{K.}~\bibnamefont{{Hasegawa}}}, \bibinfo{author}{\bibfnamefont{T.}~\bibnamefont{{Ishiyama}}}, \bibinfo{author}{\bibfnamefont{H.}~\bibnamefont{{Yajima}}}, \bibinfo{author}{\bibfnamefont{I.}~\bibnamefont{{Shimizu}}}, \bibinfo{author}{\bibfnamefont{M.}~\bibnamefont{{Umemura}}}, \bibinfo{author}{\bibfnamefont{A.}~\bibnamefont{{Konno}}}, \bibinfo{author}{\bibfnamefont{Y.}~\bibnamefont{{Harikane}}}, \bibinfo{author}{\bibfnamefont{T.}~\bibnamefont{{Shibuya}}}, \bibinfo{author}{\bibfnamefont{M.}~\bibnamefont{{Ouchi}}}, \bibnamefont{et~al.}, \bibinfo{journal}{\pasj} \textbf{\bibinfo{volume}{70}}, \bibinfo{eid}{55} (\bibinfo{year}{2018}), \eprint{1801.00067}.

\bibitem[{\citenamefont{Morales et~al.}(2021)\citenamefont{Morales, Mason, Bruton, Gronke, Haardt, and Scarlata}}]{Morales_2021}
\bibinfo{author}{\bibfnamefont{A.~M.} \bibnamefont{Morales}}, \bibinfo{author}{\bibfnamefont{C.~A.} \bibnamefont{Mason}}, \bibinfo{author}{\bibfnamefont{S.}~\bibnamefont{Bruton}}, \bibinfo{author}{\bibfnamefont{M.}~\bibnamefont{Gronke}}, \bibinfo{author}{\bibfnamefont{F.}~\bibnamefont{Haardt}}, \bibnamefont{and} \bibinfo{author}{\bibfnamefont{C.}~\bibnamefont{Scarlata}}, \bibinfo{journal}{The Astrophysical Journal} \textbf{\bibinfo{volume}{919}}, \bibinfo{pages}{120} (\bibinfo{year}{2021}), \urlprefix\url{https://dx.doi.org/10.3847/1538-4357/ac1104}.

\bibitem[{\citenamefont{Goto et~al.}(2021)\citenamefont{Goto, Shimasaku, Yamanaka, Momose, Ando, Harikane, Hashimoto, Inoue, and Ouchi}}]{Goto_2021}
\bibinfo{author}{\bibfnamefont{H.}~\bibnamefont{Goto}}, \bibinfo{author}{\bibfnamefont{K.}~\bibnamefont{Shimasaku}}, \bibinfo{author}{\bibfnamefont{S.}~\bibnamefont{Yamanaka}}, \bibinfo{author}{\bibfnamefont{R.}~\bibnamefont{Momose}}, \bibinfo{author}{\bibfnamefont{M.}~\bibnamefont{Ando}}, \bibinfo{author}{\bibfnamefont{Y.}~\bibnamefont{Harikane}}, \bibinfo{author}{\bibfnamefont{T.}~\bibnamefont{Hashimoto}}, \bibinfo{author}{\bibfnamefont{A.~K.} \bibnamefont{Inoue}}, \bibnamefont{and} \bibinfo{author}{\bibfnamefont{M.}~\bibnamefont{Ouchi}}, \bibinfo{journal}{The Astrophysical Journal} \textbf{\bibinfo{volume}{923}}, \bibinfo{pages}{229} (\bibinfo{year}{2021}), \urlprefix\url{https://dx.doi.org/10.3847/1538-4357/ac308b}.

\bibitem[{\citenamefont{Ning et~al.}(2022)\citenamefont{Ning, Jiang, Zheng, and Wu}}]{Ning_2022}
\bibinfo{author}{\bibfnamefont{Y.}~\bibnamefont{Ning}}, \bibinfo{author}{\bibfnamefont{L.}~\bibnamefont{Jiang}}, \bibinfo{author}{\bibfnamefont{Z.-Y.} \bibnamefont{Zheng}}, \bibnamefont{and} \bibinfo{author}{\bibfnamefont{J.}~\bibnamefont{Wu}}, \bibinfo{journal}{The Astrophysical Journal} \textbf{\bibinfo{volume}{926}}, \bibinfo{pages}{230} (\bibinfo{year}{2022}), \urlprefix\url{https://dx.doi.org/10.3847/1538-4357/ac4268}.

\bibitem[{\citenamefont{Umeda et~al.}(2024{\natexlab{a}})\citenamefont{Umeda, Ouchi, Kikuta, Harikane, Ono, Shibuya, Inoue, Shimasaku, Liang, Matsumoto et~al.}}]{umeda2024silverrushxivlyaluminosity}
\bibinfo{author}{\bibfnamefont{H.}~\bibnamefont{Umeda}}, \bibinfo{author}{\bibfnamefont{M.}~\bibnamefont{Ouchi}}, \bibinfo{author}{\bibfnamefont{S.}~\bibnamefont{Kikuta}}, \bibinfo{author}{\bibfnamefont{Y.}~\bibnamefont{Harikane}}, \bibinfo{author}{\bibfnamefont{Y.}~\bibnamefont{Ono}}, \bibinfo{author}{\bibfnamefont{T.}~\bibnamefont{Shibuya}}, \bibinfo{author}{\bibfnamefont{A.~K.} \bibnamefont{Inoue}}, \bibinfo{author}{\bibfnamefont{K.}~\bibnamefont{Shimasaku}}, \bibinfo{author}{\bibfnamefont{Y.}~\bibnamefont{Liang}}, \bibinfo{author}{\bibfnamefont{A.}~\bibnamefont{Matsumoto}}, \bibnamefont{et~al.} (\bibinfo{year}{2024}{\natexlab{a}}), \eprint{2411.15495}, \urlprefix\url{https://arxiv.org/abs/2411.15495}.

\bibitem[{\citenamefont{{Sobacchi} and {Mesinger}}(2015)}]{2015MNRAS.453.1843S}
\bibinfo{author}{\bibfnamefont{E.}~\bibnamefont{{Sobacchi}}} \bibnamefont{and} \bibinfo{author}{\bibfnamefont{A.}~\bibnamefont{{Mesinger}}}, \bibinfo{journal}{\mnras} \textbf{\bibinfo{volume}{453}}, \bibinfo{pages}{1843} (\bibinfo{year}{2015}), \eprint{1505.02787}.

\bibitem[{\citenamefont{{Ouchi} et~al.}(2018)\citenamefont{{Ouchi}, {Harikane}, {Shibuya}, {Shimasaku}, {Taniguchi}, {Konno}, {Kobayashi}, {Kajisawa}, {Nagao}, {Ono} et~al.}}]{2018PASJ...70S..13O}
\bibinfo{author}{\bibfnamefont{M.}~\bibnamefont{{Ouchi}}}, \bibinfo{author}{\bibfnamefont{Y.}~\bibnamefont{{Harikane}}}, \bibinfo{author}{\bibfnamefont{T.}~\bibnamefont{{Shibuya}}}, \bibinfo{author}{\bibfnamefont{K.}~\bibnamefont{{Shimasaku}}}, \bibinfo{author}{\bibfnamefont{Y.}~\bibnamefont{{Taniguchi}}}, \bibinfo{author}{\bibfnamefont{A.}~\bibnamefont{{Konno}}}, \bibinfo{author}{\bibfnamefont{M.}~\bibnamefont{{Kobayashi}}}, \bibinfo{author}{\bibfnamefont{M.}~\bibnamefont{{Kajisawa}}}, \bibinfo{author}{\bibfnamefont{T.}~\bibnamefont{{Nagao}}}, \bibinfo{author}{\bibfnamefont{Y.}~\bibnamefont{{Ono}}}, \bibnamefont{et~al.}, \bibinfo{journal}{\pasj} \textbf{\bibinfo{volume}{70}}, \bibinfo{eid}{S13} (\bibinfo{year}{2018}), \eprint{1704.07455}.

\bibitem[{\citenamefont{Umeda et~al.}(2024{\natexlab{b}})\citenamefont{Umeda, Ouchi, Nakajima, Harikane, Ono, Xu, Isobe, and Zhang}}]{Umeda_2024}
\bibinfo{author}{\bibfnamefont{H.}~\bibnamefont{Umeda}}, \bibinfo{author}{\bibfnamefont{M.}~\bibnamefont{Ouchi}}, \bibinfo{author}{\bibfnamefont{K.}~\bibnamefont{Nakajima}}, \bibinfo{author}{\bibfnamefont{Y.}~\bibnamefont{Harikane}}, \bibinfo{author}{\bibfnamefont{Y.}~\bibnamefont{Ono}}, \bibinfo{author}{\bibfnamefont{Y.}~\bibnamefont{Xu}}, \bibinfo{author}{\bibfnamefont{Y.}~\bibnamefont{Isobe}}, \bibnamefont{and} \bibinfo{author}{\bibfnamefont{Y.}~\bibnamefont{Zhang}}, \bibinfo{journal}{The Astrophysical Journal} \textbf{\bibinfo{volume}{971}}, \bibinfo{pages}{124} (\bibinfo{year}{2024}{\natexlab{b}}), \urlprefix\url{https://dx.doi.org/10.3847/1538-4357/ad554e}.

\bibitem[{\citenamefont{{Curtis-Lake} et~al.}(2023)\citenamefont{{Curtis-Lake}, {Carniani}, {Cameron}, {Charlot}, {Jakobsen}, {Maiolino}, {Bunker}, {Witstok}, {Smit}, {Chevallard} et~al.}}]{2023NatAs...7..622C}
\bibinfo{author}{\bibfnamefont{E.}~\bibnamefont{{Curtis-Lake}}}, \bibinfo{author}{\bibfnamefont{S.}~\bibnamefont{{Carniani}}}, \bibinfo{author}{\bibfnamefont{A.}~\bibnamefont{{Cameron}}}, \bibinfo{author}{\bibfnamefont{S.}~\bibnamefont{{Charlot}}}, \bibinfo{author}{\bibfnamefont{P.}~\bibnamefont{{Jakobsen}}}, \bibinfo{author}{\bibfnamefont{R.}~\bibnamefont{{Maiolino}}}, \bibinfo{author}{\bibfnamefont{A.}~\bibnamefont{{Bunker}}}, \bibinfo{author}{\bibfnamefont{J.}~\bibnamefont{{Witstok}}}, \bibinfo{author}{\bibfnamefont{R.}~\bibnamefont{{Smit}}}, \bibinfo{author}{\bibfnamefont{J.}~\bibnamefont{{Chevallard}}}, \bibnamefont{et~al.}, \bibinfo{journal}{Nature Astronomy} \textbf{\bibinfo{volume}{7}}, \bibinfo{pages}{622} (\bibinfo{year}{2023}), \eprint{2212.04568}.

\bibitem[{hsi(2024)}]{hsiao2024jwstnirspecspectroscopytriplylensed}
 (\bibinfo{year}{2024}), \eprint{2305.03042}, \urlprefix\url{https://arxiv.org/abs/2305.03042}.

\bibitem[{\citenamefont{Davies et~al.}(2018)\citenamefont{Davies, Hennawi, Bañados, Lukić, Decarli, Fan, Farina, Mazzucchelli, Rix, Venemans et~al.}}]{Davies_2018}
\bibinfo{author}{\bibfnamefont{F.~B.} \bibnamefont{Davies}}, \bibinfo{author}{\bibfnamefont{J.~F.} \bibnamefont{Hennawi}}, \bibinfo{author}{\bibfnamefont{E.}~\bibnamefont{Bañados}}, \bibinfo{author}{\bibfnamefont{Z.}~\bibnamefont{Lukić}}, \bibinfo{author}{\bibfnamefont{R.}~\bibnamefont{Decarli}}, \bibinfo{author}{\bibfnamefont{X.}~\bibnamefont{Fan}}, \bibinfo{author}{\bibfnamefont{E.~P.} \bibnamefont{Farina}}, \bibinfo{author}{\bibfnamefont{C.}~\bibnamefont{Mazzucchelli}}, \bibinfo{author}{\bibfnamefont{H.-W.} \bibnamefont{Rix}}, \bibinfo{author}{\bibfnamefont{B.~P.} \bibnamefont{Venemans}}, \bibnamefont{et~al.}, \bibinfo{journal}{The Astrophysical Journal} \textbf{\bibinfo{volume}{864}}, \bibinfo{pages}{142} (\bibinfo{year}{2018}), \urlprefix\url{https://dx.doi.org/10.3847/1538-4357/aad6dc}.

\bibitem[{\citenamefont{{Greig} et~al.}(2019)\citenamefont{{Greig}, {Mesinger}, and {Ba{\~n}ados}}}]{2019MNRAS.484.5094G}
\bibinfo{author}{\bibfnamefont{B.}~\bibnamefont{{Greig}}}, \bibinfo{author}{\bibfnamefont{A.}~\bibnamefont{{Mesinger}}}, \bibnamefont{and} \bibinfo{author}{\bibfnamefont{E.}~\bibnamefont{{Ba{\~n}ados}}}, \bibinfo{journal}{\mnras} \textbf{\bibinfo{volume}{484}}, \bibinfo{pages}{5094} (\bibinfo{year}{2019}), \eprint{1807.01593}.

\bibitem[{\citenamefont{Wang et~al.}(2020)\citenamefont{Wang, Davies, Yang, Hennawi, Fan, Barth, Jiang, Wu, Mudd, Bañados et~al.}}]{Wang_2020}
\bibinfo{author}{\bibfnamefont{F.}~\bibnamefont{Wang}}, \bibinfo{author}{\bibfnamefont{F.~B.} \bibnamefont{Davies}}, \bibinfo{author}{\bibfnamefont{J.}~\bibnamefont{Yang}}, \bibinfo{author}{\bibfnamefont{J.~F.} \bibnamefont{Hennawi}}, \bibinfo{author}{\bibfnamefont{X.}~\bibnamefont{Fan}}, \bibinfo{author}{\bibfnamefont{A.~J.} \bibnamefont{Barth}}, \bibinfo{author}{\bibfnamefont{L.}~\bibnamefont{Jiang}}, \bibinfo{author}{\bibfnamefont{X.-B.} \bibnamefont{Wu}}, \bibinfo{author}{\bibfnamefont{D.~M.} \bibnamefont{Mudd}}, \bibinfo{author}{\bibfnamefont{E.}~\bibnamefont{Bañados}}, \bibnamefont{et~al.}, \bibinfo{journal}{The Astrophysical Journal} \textbf{\bibinfo{volume}{896}}, \bibinfo{pages}{23} (\bibinfo{year}{2020}), \urlprefix\url{https://dx.doi.org/10.3847/1538-4357/ab8c45}.

\bibitem[{\citenamefont{{Totani} et~al.}(2006)\citenamefont{{Totani}, {Kawai}, {Kosugi}, {Aoki}, {Yamada}, {Iye}, {Ohta}, and {Hattori}}}]{2006PASJ...58..485T}
\bibinfo{author}{\bibfnamefont{T.}~\bibnamefont{{Totani}}}, \bibinfo{author}{\bibfnamefont{N.}~\bibnamefont{{Kawai}}}, \bibinfo{author}{\bibfnamefont{G.}~\bibnamefont{{Kosugi}}}, \bibinfo{author}{\bibfnamefont{K.}~\bibnamefont{{Aoki}}}, \bibinfo{author}{\bibfnamefont{T.}~\bibnamefont{{Yamada}}}, \bibinfo{author}{\bibfnamefont{M.}~\bibnamefont{{Iye}}}, \bibinfo{author}{\bibfnamefont{K.}~\bibnamefont{{Ohta}}}, \bibnamefont{and} \bibinfo{author}{\bibfnamefont{T.}~\bibnamefont{{Hattori}}}, \bibinfo{journal}{\pasj} \textbf{\bibinfo{volume}{58}}, \bibinfo{pages}{485} (\bibinfo{year}{2006}), \eprint{astro-ph/0512154}.

\bibitem[{\citenamefont{{Totani} et~al.}(2014)\citenamefont{{Totani}, {Aoki}, {Hattori}, {Kosugi}, {Niino}, {Hashimoto}, {Kawai}, {Ohta}, {Sakamoto}, and {Yamada}}}]{2014PASJ...66...63T}
\bibinfo{author}{\bibfnamefont{T.}~\bibnamefont{{Totani}}}, \bibinfo{author}{\bibfnamefont{K.}~\bibnamefont{{Aoki}}}, \bibinfo{author}{\bibfnamefont{T.}~\bibnamefont{{Hattori}}}, \bibinfo{author}{\bibfnamefont{G.}~\bibnamefont{{Kosugi}}}, \bibinfo{author}{\bibfnamefont{Y.}~\bibnamefont{{Niino}}}, \bibinfo{author}{\bibfnamefont{T.}~\bibnamefont{{Hashimoto}}}, \bibinfo{author}{\bibfnamefont{N.}~\bibnamefont{{Kawai}}}, \bibinfo{author}{\bibfnamefont{K.}~\bibnamefont{{Ohta}}}, \bibinfo{author}{\bibfnamefont{T.}~\bibnamefont{{Sakamoto}}}, \bibnamefont{and} \bibinfo{author}{\bibfnamefont{T.}~\bibnamefont{{Yamada}}}, \bibinfo{journal}{\pasj} \textbf{\bibinfo{volume}{66}}, \bibinfo{eid}{63} (\bibinfo{year}{2014}), \eprint{1312.3934}.

\bibitem[{\citenamefont{Hoag et~al.}(2019)\citenamefont{Hoag, Bradač, Huang, Mason, Treu, Schmidt, Trenti, Strait, Lemaux, Finney et~al.}}]{Hoag_2019}
\bibinfo{author}{\bibfnamefont{A.}~\bibnamefont{Hoag}}, \bibinfo{author}{\bibfnamefont{M.}~\bibnamefont{Bradač}}, \bibinfo{author}{\bibfnamefont{K.}~\bibnamefont{Huang}}, \bibinfo{author}{\bibfnamefont{C.}~\bibnamefont{Mason}}, \bibinfo{author}{\bibfnamefont{T.}~\bibnamefont{Treu}}, \bibinfo{author}{\bibfnamefont{K.~B.} \bibnamefont{Schmidt}}, \bibinfo{author}{\bibfnamefont{M.}~\bibnamefont{Trenti}}, \bibinfo{author}{\bibfnamefont{V.}~\bibnamefont{Strait}}, \bibinfo{author}{\bibfnamefont{B.~C.} \bibnamefont{Lemaux}}, \bibinfo{author}{\bibfnamefont{E.~Q.} \bibnamefont{Finney}}, \bibnamefont{et~al.}, \bibinfo{journal}{The Astrophysical Journal} \textbf{\bibinfo{volume}{878}}, \bibinfo{pages}{12} (\bibinfo{year}{2019}), \urlprefix\url{https://dx.doi.org/10.3847/1538-4357/ab1de7}.

\bibitem[{\citenamefont{{Planck Collaboration} et~al.}(2020{\natexlab{b}})\citenamefont{{Planck Collaboration}, {Aghanim, N.}, {Akrami, Y.}, {Ashdown, M.}, {Aumont, J.}, {Baccigalupi, C.}, {Ballardini, M.}, {Banday, A. J.}, {Barreiro, R. B.}, {Bartolo, N.} et~al.}}]{refId0}
\bibinfo{author}{\bibnamefont{{Planck Collaboration}}}, \bibinfo{author}{\bibnamefont{{Aghanim, N.}}}, \bibinfo{author}{\bibnamefont{{Akrami, Y.}}}, \bibinfo{author}{\bibnamefont{{Ashdown, M.}}}, \bibinfo{author}{\bibnamefont{{Aumont, J.}}}, \bibinfo{author}{\bibnamefont{{Baccigalupi, C.}}}, \bibinfo{author}{\bibnamefont{{Ballardini, M.}}}, \bibinfo{author}{\bibnamefont{{Banday, A. J.}}}, \bibinfo{author}{\bibnamefont{{Barreiro, R. B.}}}, \bibinfo{author}{\bibnamefont{{Bartolo, N.}}}, \bibnamefont{et~al.}, \bibinfo{journal}{A\&A} \textbf{\bibinfo{volume}{641}}, \bibinfo{pages}{A6} (\bibinfo{year}{2020}{\natexlab{b}}), \urlprefix\url{https://doi.org/10.1051/0004-6361/201833910}.

\bibitem[{\citenamefont{Fan et~al.}(2006)\citenamefont{Fan, Strauss, Becker, White, Gunn, Knapp, Richards, Schneider, Brinkmann, and Fukugita}}]{Fan_2006}
\bibinfo{author}{\bibfnamefont{X.}~\bibnamefont{Fan}}, \bibinfo{author}{\bibfnamefont{M.~A.} \bibnamefont{Strauss}}, \bibinfo{author}{\bibfnamefont{R.~H.} \bibnamefont{Becker}}, \bibinfo{author}{\bibfnamefont{R.~L.} \bibnamefont{White}}, \bibinfo{author}{\bibfnamefont{J.~E.} \bibnamefont{Gunn}}, \bibinfo{author}{\bibfnamefont{G.~R.} \bibnamefont{Knapp}}, \bibinfo{author}{\bibfnamefont{G.~T.} \bibnamefont{Richards}}, \bibinfo{author}{\bibfnamefont{D.~P.} \bibnamefont{Schneider}}, \bibinfo{author}{\bibfnamefont{J.}~\bibnamefont{Brinkmann}}, \bibnamefont{and} \bibinfo{author}{\bibfnamefont{M.}~\bibnamefont{Fukugita}}, \bibinfo{journal}{The Astronomical Journal} \textbf{\bibinfo{volume}{132}}, \bibinfo{pages}{117} (\bibinfo{year}{2006}), \urlprefix\url{https://dx.doi.org/10.1086/504836}.

\bibitem[{\citenamefont{Furlanetto et~al.}(2006)\citenamefont{Furlanetto, {Peng Oh}, and Briggs}}]{FURLANETTO2006181}
\bibinfo{author}{\bibfnamefont{S.~R.} \bibnamefont{Furlanetto}}, \bibinfo{author}{\bibfnamefont{S.}~\bibnamefont{{Peng Oh}}}, \bibnamefont{and} \bibinfo{author}{\bibfnamefont{F.~H.} \bibnamefont{Briggs}}, \bibinfo{journal}{Physics Reports} \textbf{\bibinfo{volume}{433}}, \bibinfo{pages}{181} (\bibinfo{year}{2006}), ISSN \bibinfo{issn}{0370-1573}, \urlprefix\url{https://www.sciencedirect.com/science/article/pii/S0370157306002730}.

\bibitem[{\citenamefont{{Barkana} and {Loeb}}(2008)}]{2008MNRAS.384.1069B}
\bibinfo{author}{\bibfnamefont{R.}~\bibnamefont{{Barkana}}} \bibnamefont{and} \bibinfo{author}{\bibfnamefont{A.}~\bibnamefont{{Loeb}}}, \bibinfo{journal}{\mnras} \textbf{\bibinfo{volume}{384}}, \bibinfo{pages}{1069} (\bibinfo{year}{2008}), \eprint{0705.3246}.

\bibitem[{\citenamefont{{Harker} et~al.}(2009)\citenamefont{{Harker}, {Zaroubi}, {Thomas}, {Jeli{\'c}}, {Labropoulos}, {Mellema}, {Iliev}, {Bernardi}, {Brentjens}, {de Bruyn} et~al.}}]{2009MNRAS.393.1449H}
\bibinfo{author}{\bibfnamefont{G.~J.~A.} \bibnamefont{{Harker}}}, \bibinfo{author}{\bibfnamefont{S.}~\bibnamefont{{Zaroubi}}}, \bibinfo{author}{\bibfnamefont{R.~M.} \bibnamefont{{Thomas}}}, \bibinfo{author}{\bibfnamefont{V.}~\bibnamefont{{Jeli{\'c}}}}, \bibinfo{author}{\bibfnamefont{P.}~\bibnamefont{{Labropoulos}}}, \bibinfo{author}{\bibfnamefont{G.}~\bibnamefont{{Mellema}}}, \bibinfo{author}{\bibfnamefont{I.~T.} \bibnamefont{{Iliev}}}, \bibinfo{author}{\bibfnamefont{G.}~\bibnamefont{{Bernardi}}}, \bibinfo{author}{\bibfnamefont{M.~A.} \bibnamefont{{Brentjens}}}, \bibinfo{author}{\bibfnamefont{A.~G.} \bibnamefont{{de Bruyn}}}, \bibnamefont{et~al.}, \bibinfo{journal}{\mnras} \textbf{\bibinfo{volume}{393}}, \bibinfo{pages}{1449} (\bibinfo{year}{2009}), \eprint{0809.2428}.

\bibitem[{\citenamefont{{Mesinger} and {Furlanetto}}(2007)}]{2007ApJ...669..663M}
\bibinfo{author}{\bibfnamefont{A.}~\bibnamefont{{Mesinger}}} \bibnamefont{and} \bibinfo{author}{\bibfnamefont{S.}~\bibnamefont{{Furlanetto}}}, \bibinfo{journal}{\apj} \textbf{\bibinfo{volume}{669}}, \bibinfo{pages}{663} (\bibinfo{year}{2007}), \eprint{0704.0946}.

\bibitem[{\citenamefont{{Mesinger} et~al.}(2011)\citenamefont{{Mesinger}, {Furlanetto}, and {Cen}}}]{2011MNRAS.411..955M}
\bibinfo{author}{\bibfnamefont{A.}~\bibnamefont{{Mesinger}}}, \bibinfo{author}{\bibfnamefont{S.}~\bibnamefont{{Furlanetto}}}, \bibnamefont{and} \bibinfo{author}{\bibfnamefont{R.}~\bibnamefont{{Cen}}}, \bibinfo{journal}{\mnras} \textbf{\bibinfo{volume}{411}}, \bibinfo{pages}{955} (\bibinfo{year}{2011}), \eprint{1003.3878}.

\bibitem[{\citenamefont{Park et~al.}(2019)\citenamefont{Park, Mesinger, Greig, and Gillet}}]{Park_2019}
\bibinfo{author}{\bibfnamefont{J.}~\bibnamefont{Park}}, \bibinfo{author}{\bibfnamefont{A.}~\bibnamefont{Mesinger}}, \bibinfo{author}{\bibfnamefont{B.}~\bibnamefont{Greig}}, \bibnamefont{and} \bibinfo{author}{\bibfnamefont{N.}~\bibnamefont{Gillet}}, \bibinfo{journal}{Monthly Notices of the Royal Astronomical Society} \textbf{\bibinfo{volume}{484}}, \bibinfo{pages}{933–949} (\bibinfo{year}{2019}), ISSN \bibinfo{issn}{1365-2966}, \urlprefix\url{http://dx.doi.org/10.1093/mnras/stz032}.

\bibitem[{\citenamefont{{Haslam} et~al.}(1982)\citenamefont{{Haslam}, {Salter}, {Stoffel}, and {Wilson}}}]{1982A&AS...47....1H}
\bibinfo{author}{\bibfnamefont{C.~G.~T.} \bibnamefont{{Haslam}}}, \bibinfo{author}{\bibfnamefont{C.~J.} \bibnamefont{{Salter}}}, \bibinfo{author}{\bibfnamefont{H.}~\bibnamefont{{Stoffel}}}, \bibnamefont{and} \bibinfo{author}{\bibfnamefont{W.~E.} \bibnamefont{{Wilson}}}, \bibinfo{journal}{\aaps} \textbf{\bibinfo{volume}{47}}, \bibinfo{pages}{1} (\bibinfo{year}{1982}).

\bibitem[{\citenamefont{{Coe}}(2009)}]{2009arXiv0906.4123C}
\bibinfo{author}{\bibfnamefont{D.}~\bibnamefont{{Coe}}}, \bibinfo{journal}{arXiv e-prints} \bibinfo{eid}{arXiv:0906.4123} (\bibinfo{year}{2009}), \eprint{0906.4123}.

\bibitem[{\citenamefont{Verde}(2010)}]{Verde2010}
\bibinfo{author}{\bibfnamefont{L.}~\bibnamefont{Verde}}, \emph{\bibinfo{title}{Statistical Methods in Cosmology}} (\bibinfo{publisher}{Springer Berlin Heidelberg}, \bibinfo{address}{Berlin, Heidelberg}, \bibinfo{year}{2010}), pp. \bibinfo{pages}{147--177}, ISBN \bibinfo{isbn}{978-3-642-10598-2}, \urlprefix\url{https://doi.org/10.1007/978-3-642-10598-2_4}.

\bibitem[{\citenamefont{{Prelogovi{\'c}} and {Mesinger}}(2023)}]{2023MNRAS.524.4239P}
\bibinfo{author}{\bibfnamefont{D.}~\bibnamefont{{Prelogovi{\'c}}}} \bibnamefont{and} \bibinfo{author}{\bibfnamefont{A.}~\bibnamefont{{Mesinger}}}, \bibinfo{journal}{\mnras} \textbf{\bibinfo{volume}{524}}, \bibinfo{pages}{4239} (\bibinfo{year}{2023}), \eprint{2305.03074}.

\bibitem[{\citenamefont{{Zhao} et~al.}(2024)\citenamefont{{Zhao}, {Mao}, {Zuo}, and {Wandelt}}}]{2024ApJ...973...41Z}
\bibinfo{author}{\bibfnamefont{X.}~\bibnamefont{{Zhao}}}, \bibinfo{author}{\bibfnamefont{Y.}~\bibnamefont{{Mao}}}, \bibinfo{author}{\bibfnamefont{S.}~\bibnamefont{{Zuo}}}, \bibnamefont{and} \bibinfo{author}{\bibfnamefont{B.~D.} \bibnamefont{{Wandelt}}}, \bibinfo{journal}{\apj} \textbf{\bibinfo{volume}{973}}, \bibinfo{eid}{41} (\bibinfo{year}{2024}), \eprint{2310.17602}.

\bibitem[{\citenamefont{{Sun} et~al.}(2025)\citenamefont{{Sun}, {Shao}, {Li}, {Xu}, {Wang}, and {Zhang}}}]{2025CmPhy...8..220S}
\bibinfo{author}{\bibfnamefont{T.-Y.} \bibnamefont{{Sun}}}, \bibinfo{author}{\bibfnamefont{Y.}~\bibnamefont{{Shao}}}, \bibinfo{author}{\bibfnamefont{Y.}~\bibnamefont{{Li}}}, \bibinfo{author}{\bibfnamefont{Y.}~\bibnamefont{{Xu}}}, \bibinfo{author}{\bibfnamefont{H.}~\bibnamefont{{Wang}}}, \bibnamefont{and} \bibinfo{author}{\bibfnamefont{X.}~\bibnamefont{{Zhang}}}, \bibinfo{journal}{Communications Physics} \textbf{\bibinfo{volume}{8}}, \bibinfo{eid}{220} (\bibinfo{year}{2025}), \eprint{2407.14298}.

\bibitem[{\citenamefont{{Shimabukuro} et~al.}(2014)\citenamefont{{Shimabukuro}, {Ichiki}, {Inoue}, and {Yokoyama}}}]{2014PhRvD..90h3003S}
\bibinfo{author}{\bibfnamefont{H.}~\bibnamefont{{Shimabukuro}}}, \bibinfo{author}{\bibfnamefont{K.}~\bibnamefont{{Ichiki}}}, \bibinfo{author}{\bibfnamefont{S.}~\bibnamefont{{Inoue}}}, \bibnamefont{and} \bibinfo{author}{\bibfnamefont{S.}~\bibnamefont{{Yokoyama}}}, \bibinfo{journal}{\prd} \textbf{\bibinfo{volume}{90}}, \bibinfo{eid}{083003} (\bibinfo{year}{2014}), \eprint{1403.1605}.

\bibitem[{\citenamefont{{Shimabukuro} et~al.}(2020{\natexlab{a}})\citenamefont{{Shimabukuro}, {Ichiki}, and {Kadota}}}]{2020PhRvD.101d3516S}
\bibinfo{author}{\bibfnamefont{H.}~\bibnamefont{{Shimabukuro}}}, \bibinfo{author}{\bibfnamefont{K.}~\bibnamefont{{Ichiki}}}, \bibnamefont{and} \bibinfo{author}{\bibfnamefont{K.}~\bibnamefont{{Kadota}}}, \bibinfo{journal}{\prd} \textbf{\bibinfo{volume}{101}}, \bibinfo{eid}{043516} (\bibinfo{year}{2020}{\natexlab{a}}), \eprint{1910.06011}.

\bibitem[{\citenamefont{{Shimabukuro} et~al.}(2020{\natexlab{b}})\citenamefont{{Shimabukuro}, {Ichiki}, and {Kadota}}}]{2020PhRvD.102b3522S}
\bibinfo{author}{\bibfnamefont{H.}~\bibnamefont{{Shimabukuro}}}, \bibinfo{author}{\bibfnamefont{K.}~\bibnamefont{{Ichiki}}}, \bibnamefont{and} \bibinfo{author}{\bibfnamefont{K.}~\bibnamefont{{Kadota}}}, \bibinfo{journal}{\prd} \textbf{\bibinfo{volume}{102}}, \bibinfo{eid}{023522} (\bibinfo{year}{2020}{\natexlab{b}}), \eprint{2005.05589}.

\bibitem[{\citenamefont{{Shimabukuro} et~al.}(2023{\natexlab{b}})\citenamefont{{Shimabukuro}, {Ichiki}, and {Kadota}}}]{2023PhRvD.107l3520S}
\bibinfo{author}{\bibfnamefont{H.}~\bibnamefont{{Shimabukuro}}}, \bibinfo{author}{\bibfnamefont{K.}~\bibnamefont{{Ichiki}}}, \bibnamefont{and} \bibinfo{author}{\bibfnamefont{K.}~\bibnamefont{{Kadota}}}, \bibinfo{journal}{\prd} \textbf{\bibinfo{volume}{107}}, \bibinfo{eid}{123520} (\bibinfo{year}{2023}{\natexlab{b}}), \eprint{2212.08409}.

\bibitem[{\citenamefont{{Shao} et~al.}(2023)\citenamefont{{Shao}, {Xu}, {Wang}, {Yang}, {Li}, {Zhang}, and {Chen}}}]{2023NatAs...7.1116S}
\bibinfo{author}{\bibfnamefont{Y.}~\bibnamefont{{Shao}}}, \bibinfo{author}{\bibfnamefont{Y.}~\bibnamefont{{Xu}}}, \bibinfo{author}{\bibfnamefont{Y.}~\bibnamefont{{Wang}}}, \bibinfo{author}{\bibfnamefont{W.}~\bibnamefont{{Yang}}}, \bibinfo{author}{\bibfnamefont{R.}~\bibnamefont{{Li}}}, \bibinfo{author}{\bibfnamefont{X.}~\bibnamefont{{Zhang}}}, \bibnamefont{and} \bibinfo{author}{\bibfnamefont{X.}~\bibnamefont{{Chen}}}, \bibinfo{journal}{Nature Astronomy} \textbf{\bibinfo{volume}{7}}, \bibinfo{pages}{1116} (\bibinfo{year}{2023}), \eprint{2307.04130}.

\bibitem[{\citenamefont{{Shimabukuro} et~al.}(2025)\citenamefont{{Shimabukuro}, {Xu}, and {Shao}}}]{2025arXiv250414656S}
\bibinfo{author}{\bibfnamefont{H.}~\bibnamefont{{Shimabukuro}}}, \bibinfo{author}{\bibfnamefont{Y.}~\bibnamefont{{Xu}}}, \bibnamefont{and} \bibinfo{author}{\bibfnamefont{Y.}~\bibnamefont{{Shao}}}, \bibinfo{journal}{arXiv e-prints} \bibinfo{eid}{arXiv:2504.14656} (\bibinfo{year}{2025}), \eprint{2504.14656}.

\bibitem[{\citenamefont{{{\v{S}}oltinsk{\'y}} et~al.}(2025)\citenamefont{{{\v{S}}oltinsk{\'y}}, {Kulkarni}, {Tendulkar}, and {Bolton}}}]{2025MNRAS.537..364S}
\bibinfo{author}{\bibfnamefont{T.}~\bibnamefont{{{\v{S}}oltinsk{\'y}}}}, \bibinfo{author}{\bibfnamefont{G.}~\bibnamefont{{Kulkarni}}}, \bibinfo{author}{\bibfnamefont{S.~P.} \bibnamefont{{Tendulkar}}}, \bibnamefont{and} \bibinfo{author}{\bibfnamefont{J.~S.} \bibnamefont{{Bolton}}}, \bibinfo{journal}{\mnras} \textbf{\bibinfo{volume}{537}}, \bibinfo{pages}{364} (\bibinfo{year}{2025}), \eprint{2412.06879}.

\bibitem[{\citenamefont{{Shao} et~al.}(2025)\citenamefont{{Shao}, {Du}, {Li}, and {Zhang}}}]{2025PhLB..86239342S}
\bibinfo{author}{\bibfnamefont{Y.}~\bibnamefont{{Shao}}}, \bibinfo{author}{\bibfnamefont{G.-H.} \bibnamefont{{Du}}}, \bibinfo{author}{\bibfnamefont{T.-N.} \bibnamefont{{Li}}}, \bibnamefont{and} \bibinfo{author}{\bibfnamefont{X.}~\bibnamefont{{Zhang}}}, \bibinfo{journal}{Physics Letters B} \textbf{\bibinfo{volume}{862}}, \bibinfo{eid}{139342} (\bibinfo{year}{2025}), \eprint{2501.00769}.

\bibitem[{\citenamefont{{Liu} and {Tegmark}}(2011)}]{2011PhRvD..83j3006L}
\bibinfo{author}{\bibfnamefont{A.}~\bibnamefont{{Liu}}} \bibnamefont{and} \bibinfo{author}{\bibfnamefont{M.}~\bibnamefont{{Tegmark}}}, \bibinfo{journal}{\prd} \textbf{\bibinfo{volume}{83}}, \bibinfo{eid}{103006} (\bibinfo{year}{2011}), \eprint{1103.0281}.

\bibitem[{\citenamefont{{Petrovic} and {Oh}}(2011)}]{2011MNRAS.413.2103P}
\bibinfo{author}{\bibfnamefont{N.}~\bibnamefont{{Petrovic}}} \bibnamefont{and} \bibinfo{author}{\bibfnamefont{S.~P.} \bibnamefont{{Oh}}}, \bibinfo{journal}{\mnras} \textbf{\bibinfo{volume}{413}}, \bibinfo{pages}{2103} (\bibinfo{year}{2011}), \eprint{1010.4109}.

\bibitem[{\citenamefont{{Majumdar} et~al.}(2018)\citenamefont{{Majumdar}, {Pritchard}, {Mondal}, {Watkinson}, {Bharadwaj}, and {Mellema}}}]{2018MNRAS.476.4007M}
\bibinfo{author}{\bibfnamefont{S.}~\bibnamefont{{Majumdar}}}, \bibinfo{author}{\bibfnamefont{J.~R.} \bibnamefont{{Pritchard}}}, \bibinfo{author}{\bibfnamefont{R.}~\bibnamefont{{Mondal}}}, \bibinfo{author}{\bibfnamefont{C.~A.} \bibnamefont{{Watkinson}}}, \bibinfo{author}{\bibfnamefont{S.}~\bibnamefont{{Bharadwaj}}}, \bibnamefont{and} \bibinfo{author}{\bibfnamefont{G.}~\bibnamefont{{Mellema}}}, \bibinfo{journal}{\mnras} \textbf{\bibinfo{volume}{476}}, \bibinfo{pages}{4007} (\bibinfo{year}{2018}), \eprint{1708.08458}.

\bibitem[{\citenamefont{{Watkinson} et~al.}(2019)\citenamefont{{Watkinson}, {Giri}, {Ross}, {Dixon}, {Iliev}, {Mellema}, and {Pritchard}}}]{2019MNRAS.482.2653W}
\bibinfo{author}{\bibfnamefont{C.~A.} \bibnamefont{{Watkinson}}}, \bibinfo{author}{\bibfnamefont{S.~K.} \bibnamefont{{Giri}}}, \bibinfo{author}{\bibfnamefont{H.~E.} \bibnamefont{{Ross}}}, \bibinfo{author}{\bibfnamefont{K.~L.} \bibnamefont{{Dixon}}}, \bibinfo{author}{\bibfnamefont{I.~T.} \bibnamefont{{Iliev}}}, \bibinfo{author}{\bibfnamefont{G.}~\bibnamefont{{Mellema}}}, \bibnamefont{and} \bibinfo{author}{\bibfnamefont{J.~R.} \bibnamefont{{Pritchard}}}, \bibinfo{journal}{\mnras} \textbf{\bibinfo{volume}{482}}, \bibinfo{pages}{2653} (\bibinfo{year}{2019}), \eprint{1808.02372}.

\bibitem[{\citenamefont{{Hutter} et~al.}(2020)\citenamefont{{Hutter}, {Watkinson}, {Seiler}, {Dayal}, {Sinha}, and {Croton}}}]{2020MNRAS.492..653H}
\bibinfo{author}{\bibfnamefont{A.}~\bibnamefont{{Hutter}}}, \bibinfo{author}{\bibfnamefont{C.~A.} \bibnamefont{{Watkinson}}}, \bibinfo{author}{\bibfnamefont{J.}~\bibnamefont{{Seiler}}}, \bibinfo{author}{\bibfnamefont{P.}~\bibnamefont{{Dayal}}}, \bibinfo{author}{\bibfnamefont{M.}~\bibnamefont{{Sinha}}}, \bibnamefont{and} \bibinfo{author}{\bibfnamefont{D.~J.} \bibnamefont{{Croton}}}, \bibinfo{journal}{\mnras} \textbf{\bibinfo{volume}{492}}, \bibinfo{pages}{653} (\bibinfo{year}{2020}), \eprint{1907.04342}.

\bibitem[{\citenamefont{{Yoshiura} et~al.}(2017)\citenamefont{{Yoshiura}, {Shimabukuro}, {Takahashi}, and {Matsubara}}}]{2017MNRAS.465..394Y}
\bibinfo{author}{\bibfnamefont{S.}~\bibnamefont{{Yoshiura}}}, \bibinfo{author}{\bibfnamefont{H.}~\bibnamefont{{Shimabukuro}}}, \bibinfo{author}{\bibfnamefont{K.}~\bibnamefont{{Takahashi}}}, \bibnamefont{and} \bibinfo{author}{\bibfnamefont{T.}~\bibnamefont{{Matsubara}}}, \bibinfo{journal}{\mnras} \textbf{\bibinfo{volume}{465}}, \bibinfo{pages}{394} (\bibinfo{year}{2017}), \eprint{1602.02351}.

\bibitem[{\citenamefont{{Chen} et~al.}(2019)\citenamefont{{Chen}, {Xu}, {Wang}, and {Chen}}}]{2019ApJ...885...23C}
\bibinfo{author}{\bibfnamefont{Z.}~\bibnamefont{{Chen}}}, \bibinfo{author}{\bibfnamefont{Y.}~\bibnamefont{{Xu}}}, \bibinfo{author}{\bibfnamefont{Y.}~\bibnamefont{{Wang}}}, \bibnamefont{and} \bibinfo{author}{\bibfnamefont{X.}~\bibnamefont{{Chen}}}, \bibinfo{journal}{\apj} \textbf{\bibinfo{volume}{885}}, \bibinfo{eid}{23} (\bibinfo{year}{2019}), \eprint{1812.10333}.

\bibitem[{\citenamefont{{Giri} and {Mellema}}(2021)}]{2021MNRAS.505.1863G}
\bibinfo{author}{\bibfnamefont{S.~K.} \bibnamefont{{Giri}}} \bibnamefont{and} \bibinfo{author}{\bibfnamefont{G.}~\bibnamefont{{Mellema}}}, \bibinfo{journal}{\mnras} \textbf{\bibinfo{volume}{505}}, \bibinfo{pages}{1863} (\bibinfo{year}{2021}), \eprint{2012.12908}.

\bibitem[{\citenamefont{{Zhao} et~al.}(2022)\citenamefont{{Zhao}, {Mao}, {Cheng}, and {Wandelt}}}]{2022ApJ...926..151Z}
\bibinfo{author}{\bibfnamefont{X.}~\bibnamefont{{Zhao}}}, \bibinfo{author}{\bibfnamefont{Y.}~\bibnamefont{{Mao}}}, \bibinfo{author}{\bibfnamefont{C.}~\bibnamefont{{Cheng}}}, \bibnamefont{and} \bibinfo{author}{\bibfnamefont{B.~D.} \bibnamefont{{Wandelt}}}, \bibinfo{journal}{\apj} \textbf{\bibinfo{volume}{926}}, \bibinfo{eid}{151} (\bibinfo{year}{2022}), \eprint{2105.03344}.

\bibitem[{\citenamefont{{Gillet} et~al.}(2019)\citenamefont{{Gillet}, {Mesinger}, {Greig}, {Liu}, and {Ucci}}}]{2019MNRAS.484..282G}
\bibinfo{author}{\bibfnamefont{N.}~\bibnamefont{{Gillet}}}, \bibinfo{author}{\bibfnamefont{A.}~\bibnamefont{{Mesinger}}}, \bibinfo{author}{\bibfnamefont{B.}~\bibnamefont{{Greig}}}, \bibinfo{author}{\bibfnamefont{A.}~\bibnamefont{{Liu}}}, \bibnamefont{and} \bibinfo{author}{\bibfnamefont{G.}~\bibnamefont{{Ucci}}}, \bibinfo{journal}{\mnras} \textbf{\bibinfo{volume}{484}}, \bibinfo{pages}{282} (\bibinfo{year}{2019}), \eprint{1805.02699}.

\bibitem[{\citenamefont{{Furlanetto} et~al.}(2006)\citenamefont{{Furlanetto}, {Oh}, and {Briggs}}}]{2006PhR...433..181F}
\bibinfo{author}{\bibfnamefont{S.~R.} \bibnamefont{{Furlanetto}}}, \bibinfo{author}{\bibfnamefont{S.~P.} \bibnamefont{{Oh}}}, \bibnamefont{and} \bibinfo{author}{\bibfnamefont{F.~H.} \bibnamefont{{Briggs}}}, \bibinfo{journal}{\physrep} \textbf{\bibinfo{volume}{433}}, \bibinfo{pages}{181} (\bibinfo{year}{2006}), \eprint{astro-ph/0608032}.

\bibitem[{\citenamefont{{Clark} et~al.}(2018)\citenamefont{{Clark}, {Dutta}, {Gao}, {Ma}, and {Strigari}}}]{2018PhRvD..98d3006C}
\bibinfo{author}{\bibfnamefont{S.~J.} \bibnamefont{{Clark}}}, \bibinfo{author}{\bibfnamefont{B.}~\bibnamefont{{Dutta}}}, \bibinfo{author}{\bibfnamefont{Y.}~\bibnamefont{{Gao}}}, \bibinfo{author}{\bibfnamefont{Y.-Z.} \bibnamefont{{Ma}}}, \bibnamefont{and} \bibinfo{author}{\bibfnamefont{L.~E.} \bibnamefont{{Strigari}}}, \bibinfo{journal}{\prd} \textbf{\bibinfo{volume}{98}}, \bibinfo{eid}{043006} (\bibinfo{year}{2018}), \eprint{1803.09390}.

\bibitem[{\citenamefont{{Mu{\~n}oz} et~al.}(2015)\citenamefont{{Mu{\~n}oz}, {Ali-Ha{\"\i}moud}, and {Kamionkowski}}}]{2015PhRvD..92h3508M}
\bibinfo{author}{\bibfnamefont{J.~B.} \bibnamefont{{Mu{\~n}oz}}}, \bibinfo{author}{\bibfnamefont{Y.}~\bibnamefont{{Ali-Ha{\"\i}moud}}}, \bibnamefont{and} \bibinfo{author}{\bibfnamefont{M.}~\bibnamefont{{Kamionkowski}}}, \bibinfo{journal}{\prd} \textbf{\bibinfo{volume}{92}}, \bibinfo{eid}{083508} (\bibinfo{year}{2015}), \eprint{1506.04152}.

\end{thebibliography}
\endgroup


\end{document}